\documentclass[prx,aps,showkeys,showpacs,footnoteinbib,onecolumn,
unsortedaddress]{revtex4-2}
\usepackage{graphicx} 
\usepackage{dcolumn} 
\usepackage{amssymb}

\usepackage{physics}
\usepackage{natbib}
\usepackage{xcolor}
\usepackage{soul}
\usepackage{siunitx}
\usepackage{float}
\usepackage{esdiff}
\usepackage{bbold}
\usepackage{mathtools}
\usepackage{todonotes}
\usepackage{hyperref}
\usepackage{amsmath,amssymb,epic,eepic}
\usepackage[binary-units]{siunitx}
\usepackage[utf8]{inputenc}
\usepackage{graphicx}
\usepackage{hyperref}
\usepackage[capitalise]{cleveref}
\usepackage{todonotes}
\usepackage{bbm}
\usepackage{braket}
\usepackage{fontawesome}
\usepackage{ccaption}

\newcommand{\cG}{\mathcal{G}}

\usepackage[labelfont=bf]{caption}

\begin{document}

\title{Time-domain braiding of anyons}

\author{ M. Ruelle$^{1}$, E. Frigerio$^{1}$,   E. Baudin$^{1 }$, J.-M. Berroir$^{1 }$,   B.
Pla\c{c}ais$^{1 }$,  B. Gr\'{e}maud $^{2}$,\\
T. Jonckheere$^{2}$, T. Martin $^{2}$, J. Rech$^{2}$,  A. Cavanna$^{3}$, U. Gennser$^{3}$, Y. Jin$^{3}$, \\
 G. M\'{e}nard$^{1}$, and G. F\`{e}ve$^{1 \ast}$ \\
\normalsize{$^{1}$ Laboratoire de Physique de l\textquoteright Ecole normale sup\'erieure, ENS, Universit\'e
PSL,  }\\
\normalsize{CNRS, Sorbonne Universit\'e, Universit\'e Paris Cit\'e, F-75005 Paris, France}\\
\normalsize{$^{2}$ Aix Marseille Univ, Universit\'{e} de Toulon, CNRS, CPT, Marseille, France. }\\
\normalsize{$^{3}$ Centre de Nanosciences et de Nanotechnologies (C2N), CNRS, }\\
\normalsize{ Universit\'{e} Paris-Saclay, 91120 Palaiseau, France. }\\
\normalsize{$^\ast$ To whom correspondence should be addressed; E-mail:  gwendal.feve@ens.fr.}\\}

\begin{abstract}
Contrary to fermions and bosons, anyons are quasiparticles that keep a robust memory of particle exchanges via a braiding phase factor. This provides them with unique dynamical properties so far unexplored. When an anyon excitation is emitted toward a quantum point contact (QPC) in a fractional quantum Hall (FQH) fluid, this memory translates into tunneling events that may occur long after the anyon excitation has exited the QPC. Here, we use triggered anyon pulses incident on a QPC in a $\nu= 1/3$ FQH fluid to investigate anyon tunneling in the time domain. We observe that braiding increases the tunneling timescale, which is set by the temperature and the anyon scaling dimension that characterizes the edge state dynamics. This contrasts with the electron behavior where braiding is absent and the tunneling timescale is set by the temporal width of the generated electron pulses. Our experiment introduces time-domain measurements for characterizing the braiding phase and scaling dimension of anyons.
\end{abstract}

\maketitle

\section*{Introduction}

Two-dimensional systems can host anyons \cite{Leinaas1977,Goldin1980,Wilczek1982,Wilczek1982b}, topological excitations obeying fractional statistics in between fermions and bosons. In the simplest case of abelian anyons, fractional statistics implies that  moving one anyon around another (a braiding operation \cite{Note0}) results in the accumulation of a non-trivial phase factor $e^{i \theta} \neq 1$. This robust memory of braiding operations is at the heart of topological quantum computing, where unitary transformations are performed by braiding non-abelian anyons \cite{Kitaev2003,Nayak2008}, for which also the order of exchanging the particles plays a role. Fractional quantum Hall conductors \cite{Tsui1982} (FQHC) have been suggested as systems of interest hosting anyons as the elementary excitations of their insulating bulk  \cite{Arovas1984,Halperin1984}. However, probing the bulk is experimentally challenging, and most characterizations of FQHC have been performed via transport measurements along their conducting edge channels.  By weakly coupling two counter-propagating edge channels with a quantum point contact (QPC),  one can randomly transfer anyon excitations from one edge to the other. As they are transferred through the insulating FQH fluid at the narrow gap of the QPC, these excitations inherit the topological properties of the bulk with a quantized fractional charge $e^*$ and braiding phase $\theta$.

More than twenty years ago, single QPCs were used for the determination of the fractional charge  $e^*$ \cite{Kane1994,Saminadayar1997,dePicciotto1997}. Regarding anyon fractional statistics, more complex experiments relying on single \cite{Chamon1997,Law2006,Halperin2011} and two particle \cite{Safi2001,Vishveshwara2003,Kim2005,Campagnano2012,Rosenow2016} interferometry, and using sample geometries comprising several QPCs have been suggested. Following these proposals, fractional statistics of anyons were experimentally demonstrated in 2020 at the filling factor $\nu=1/3$ in collider \cite{Bartolomei2020} and Fabry-Perot geometries \cite{Nakamura2020}, subsequently followed by several experiments \cite{Ruelle2023,Glidic2023,Lee2023,Nakamura2023,Willett2023,Heiblum2023} confirming the previous results and extending them to other filling factors. Apart from $e^*$ and $\theta$, anyon dynamics at the edge of a FQHC are also characterized by a third parameter, the scaling dimension $\delta$ \cite{Wen1991, Schiller2022}, which sets the characteristic time of the temporal correlations of edge anyonic excitations. Unlike the fractional charge and statistics that are imposed by the topological properties of the bulk, $\delta$ only reflects the properties of the edge. Therefore, it is not necessarily quantized and may depend on local parameters such as the interaction strength and the smoothness of the edge \cite{Rosenow2002}. Experimental attempts at extracting $\delta$ have relied on the measurement of the non-linear $I-V$ characteristics resulting from anyon tunneling at a QPC \cite{Roddaro2004,Radu2008,Baer2014}. Despite qualitative similarities with theory predictions within the Luttinger model \cite{Wen1991}, no quantitative agreement was observed, hampering any quantitative estimate of  $\delta$ \cite{Note1}.

Almost all previous experiments investigating anyon properties have focused on the DC regime using stationary anyon currents. As such, they do not offer the possibility to trigger the emission of anyons in the circuit in order to probe their dynamics. As with all elementary excitations, controlling anyon emission at the single-particle level is crucial for the development of important experiments, such as probing anyon braiding in the time-domain, performing on-demand braiding operations, and accessing the characteristic timescales of anyon transfer in a circuit. Furthermore, by bringing additional information with respect to their DC counterpart, time-resolved experiments provide a way to disentangle the respective roles of the different properties of anyons, such as their fractional charge, statistics and scaling dimension.  However, emitting single anyons on demand is challenging. Deterministic electron emission has been realized in the integer quantum Hall regime by using a driven quantum dot \cite{Feve2007}, but these methods cannot be applied to the anyon case. There is a mutual incompatibility  between the confinement in a quantum dot and the emission of anyons, as the necessary barrier favors the tunneling of electrons over anyons. There have been suggestions to circumvent this problem by using antidots\cite{Ferraro2015}, but a simpler route was recently proposed. It consists in applying fast time-dependent drives to a one-dimensional chiral edge channel\cite{Lee2020,Mora2022,Jonckheere2023} via a metallic gate or an ohmic contact. Due to their gapless and linear low energy spectrum, the excitations of chiral edge channels consist in continuous deformations of the edge charge density with a chiral propagation\cite{Wen1995}.  By driving a one-dimensional edge channel with a short time voltage pulse, one can thus generate a propagating current pulse carrying a total charge $q = \int_{-\infty}^{+\infty} I(x,t) dt$ that varies continuously with the amplitude (and width) of the applied drive. This implies that, unlike a tunnel process, the number of anyons carried per pulse $N=q/e^*$  is not quantized to a discrete number but can be controlled with the drive amplitude, offering a knob for tuning anyon braiding in time-resolved experiments.

In this work, we investigate the properties of anyons tunneling at a QPC in a $\nu=1/3$ FQHC: their braiding phase $\theta$ characterizing their exchange properties and their scaling dimension $\delta$ characterizing their dynamical properties. To achieve this, we generate short time anyon current pulses characterized by their temporal width $W$ (here fixed to $\approx 100$ ps) and by the number of anyons $N$ carried by each pulse. We investigate anyon braiding by studying the partitioning of the generated anyon pulses at a QPC. When anyons are impinging on a QPC, the dominant mechanism for particle transfer is not the direct tunneling of the incoming excitations, but rather a braiding process between the incoming excitations and particle-hole excitations created at the QPC\cite{Morel2022,Lee2022,Mora2022,Schiller2023}. Anyon tunneling is then governed by the mutual braiding phase $\theta \times N$ \cite{Supp} between the generated anyon pulses and the topological anyons tunneling at the QPC. In contrast to previous experiments, the mutual braiding phase can be varied by tuning the dimensionless charge $N$ carried by each anyon pulse. Additionally, the triggering of anyon emission allows us to investigate anyon braiding in the time-domain and to characterize its effect on the dynamics of anyon tunneling at the QPC. We observe that for a non-trivial braiding ($\theta\times N<2\pi$), anyons keep a memory of braiding processes occurring at the QPC, and tunneling may happen long after the emitted anyon pulse has exited the QPC.   The characteristic timescale for anyon tunneling is then set by the temporal decay of the anyon correlation function over a time $\tau_{\delta}=\hbar/(\pi k_B T_{\text{el}} \delta)$, parameterized by the scaling dimension  $\delta $. It can be understood as the characteristic time on which the anyon memory is erased at the edge of the FQH fluid. It implies that the effect of braiding on anyon tunneling dynamics is more important at low electronic temperature $T_{\text{el}}$ and for small values of $\delta$. By measuring the dependence of $\tau_{\delta}$ with the temperature $T_{\text{el}}$, we extract  $\delta =0.66 \pm 0.08$ for a wide range of QPC parameters. It differs from the naive expectation $\delta= \nu =1/3$ for a Laughlin state \cite{Wen1991}, showing that $\delta$ is indeed non-universal. In contrast, when braiding is trivial ($\theta\times N=2\pi$), we observe that the characteristic tunneling timescale is set by the temporal width of the generated pulses, as one would naively expect. For pulses of narrow width $W\approx 100$ ps, this is observed as a reduction by a factor 2 of the characteristic tunneling time when braiding becomes trivial. Finally, by measuring that the suppression of braiding effects occurs for pulses containing $N= 3$ anyons, we confirm the quantized value of the braiding phase of topological anyons, $\theta=2\pi/3$ at filling factor $1/3$. By giving immediate access to the scaling dimension $\delta$ and to the anyon braiding phase $\theta$, this experiment strikingly illustrates the relevance of triggered anyon emission, allowing us to explore the consequences of fractional statistics. These results will be instrumental for the characterization of topological anyon excitations in more complex FQHCs, where the fractional charge and the statistics are not characterized by the same quantized number.

\section*{Braiding at a QPC and anyon pulses}

Anyon tunneling between counterpropagating edge channels can be induced by using a QPC. In the limit of weak backscattering, only two tunneling processes, represented on Fig.1a in the equilibrium situation, need to be considered. The left panel represents the tunneling at time $t'$ of an anyon (in dark blue) from channel 1 to channel 2, leaving a hole (in light blue) in channel 1. This tunnel process can be represented by the quantum state $| t' \rangle_+$, where the sign $+$ stands for the direction of tunneling (from channel 1 to channel 2). The associated tunneling rate $\Gamma_+(t)$ results from the time-domain interference \cite{Morel2022,Lee2022,Mora2022,Schiller2023} (see also \cite{Supp}) between tunneling events occurring at two different times $t$ and $t'$, and can be written as the following quantum overlap:  $\Gamma_+ (t) \propto \Re[\int dt' \langle t | t' \rangle_+ ]$. Together with the forward tunneling $| t' \rangle_+$, one needs also to consider backward tunneling events $| t' \rangle_-$ describing the transfer of an anyon at time $t'$ from channel 2 to channel 1 (see Fig.1a, right panel), which occur with a rate $\Gamma_- (t) \propto \Re[\int dt' \langle t | t' \rangle_- ]$. At equilibrium, where no anyon sources are connected at the input of the QPC, the two tunneling rates are equal, $\Gamma_+ (t)=\Gamma_- (t)$, and describe random transfers of anyons from one edge to the other. The magnitude of the equilibrium tunneling rates is then related to the equilibrium correlation function of the fractional quantum Hall fluid: $\langle t| t' \rangle_+=\langle t| t' \rangle_-=\mathcal{G}_{\delta}^2(t-t')$, which decays on long times with the characteristic timescale $\tau_{\delta}$. The circumstances are completely different when considering the non-equilibrium case where an anyon source is connected at input 1 of the QPC. Fig.1b represents the forward tunneling process occurring with rate $\Gamma_+(t)$ in the simple case where a single anyon excitation (in red on Fig.1b) is emitted by the source and reaches the QPC at time $t_0$. In this case,  $\Gamma_+(t)$ is governed by the braiding processes occurring between the red anyon excitation emitted by the source and the anyon quasihole excitation  created in channel 1 by the anyon tunneling process. This braiding mechanism is illustrated on Fig.1b, where the left panel (respectively central panel) represents the quantum state  $|t t_0 \rangle_+$ (resp. $|t' t_0 \rangle_+$) describing forward anyon tunneling (from channel 1 to channel 2) at time $t$ (resp. $t'$) in the presence of a red anyon reaching the QPC at time $t_0$. For $t' \leq t_0 \leq t$ the red anyon crosses the QPC after the tunneling event at time $t$ (left panel) but before the tunneling event at time $t'$ (central panel). Anyon braiding thus emerges from the time-domain interferometry\cite{Morel2022,Lee2022,Mora2022,Schiller2023} (see also \cite{Supp}) between the two tunneling processes $|t t_0 \rangle_+$ and $|t' t_0 \rangle_+$. As illustrated on Fig.1b (right panel), the overlap $\langle t_0 t |t' t_0 \rangle_+$ differs from the equilibrium value by a relative phase given by the mutual braiding phase between the blue anyon quasihole and the red anyons having crossed the QPC in the time window $t-t'$. 

\begin{figure}
\begin{center}
\includegraphics[width=1
\columnwidth,keepaspectratio]{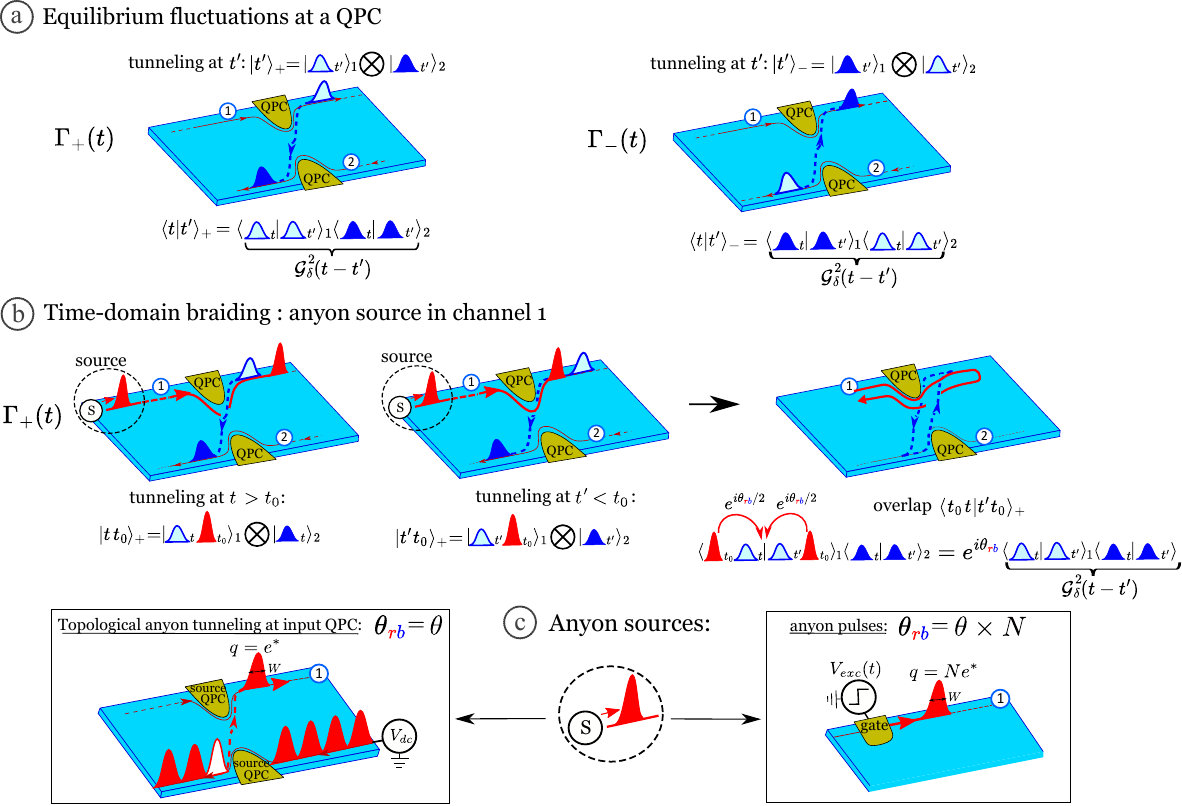}
\caption{ \small{\textbf{Anyon braiding at a QPC.} (\textbf{a}) Anyon tunneling at a QPC, equilibrium case. Left: forward tunneling process described by the state $|t' \rangle_+$,  where an anyon (in dark blue) tunnels from channel 1 to channel 2 at time $t'$, leaving a hole (in light blue) in channel 1. The forward tunneling rate $\Gamma_+(t)$ results from the time-domain interference between the different tunneling processes $|t \rangle_+$ and $|t' \rangle_+$ occurring at different times, and can be computed from the overlap:  $\Gamma_+(t)=  \Re [\int dt'  \langle t |t' \rangle_+]$. Right: backward tunneling process described by the state $|t' \rangle_-$, where an anyon tunnels from channel 2 to channel 1 at time $t'$. At equilibrium,  $\Gamma_+(t)= \Gamma_-(t)$ and the tunneling current $I_T(t)=e^* [\Gamma_+(t)- \Gamma_-(t)]$ vanishes. (\textbf{b}) Anyon braiding when an anyon source is connected at input 1.  In the presence of an anyon excitation (in red) incoming on the QPC at time $t_0$, the processes $|t t_0 \rangle_+$ (left panel) and $|t' t_0 \rangle_+$ (central panel) represent the tunneling of the blue anyon after (for $t\geq t_0$) and before (for $t' \leq t_0$) the red anyon has crossed the QPC. The overlap  $\langle t_0 t|t' t_0 \rangle_+$ is sketched on the right panel, where a reversed time propagation is used to represent the ket $\langle t_0 t|$. As can be seen on the sketch (see \cite{Supp} for a more detailed explanation)  $\langle t_0 t|t' t_0 \rangle_+$ differs from the overlap at equilibrium $\langle t|t' \rangle_+$ by the braiding phase factor $e^{i \theta_{rb}}$, where $ \theta_{rb}$ is the mutual braiding phase between the red and blue anyons. 
 (\textbf{c}) Two different sources of anyons may be used. In anyon collider experiments (left panel), a source QPC biased by a DC voltage generates a dilute and stationary beam of anyons from anyon tunneling. $ \theta_{rb}$ thus takes the quantized value $\theta$ imposed by the bulk topological order. In this manuscript, we investigate the triggered emission of an anyon pulse with charge $q=Ne^*$ from a voltage driven metallic top gate (right panel). The mutual braiding then equals $ \theta_{rb}=\theta \times N$, where $N$  is the number of anyons carried per pulse.   } }
 \end{center}
\end{figure}

This braiding phase thus depends on the nature of the blue and red quasiparticles. The blue anyons tunneling at the QPC are the topological excitations whose properties are imposed by the bulk topological order, while the nature of the red anyons depends on the anyon source. In anyon collider experiments \cite{Bartolomei2020,Ruelle2023,Glidic2023,Lee2023}, the red anyons are also randomly generated by a tunneling process at a source QPC driven by a dc voltage (see Fig.1c) and are thus also topological anyons. The mutual braiding phase is thus quantized to the value $\theta$, which has proved decisive for the study of anyon fractional statistics in collider experiments.  We investigate here the emission of current pulses via a metallic gate (represented on the right panel of Fig.1c), each pulse carrying a charge $q=Ne^*$.  As a result, the mutual braiding phase between the blue and red anyons is $\theta \times N$ \cite{Supp}.  When $N$ is an integer, it can be straightforwardly understood from the fusion of $N$ anyons\cite{Nayak2008}. Interestingly, the gapless spectrum of edge channels allows us to generalize this expression to non-integer values of $N$, so that the braiding phase can be continuously tuned in experiments.

In the case of the backward tunneling rate $\Gamma_-(t)$, discussed in more detail in\cite{Supp}, the anyon tunneling process occurs in the opposite direction (from channel 2 to channel 1). The red anyon braids with an anyon quasiparticle instead of an anyon quasihole, resulting in an opposite braiding phase compared to the forward tunneling case. The time dependent tunneling rates $\Gamma_{+}(t)$ and $\Gamma_{- }(t)$, as well as the tunneling current through the QPC $I_T(t)=e^* [\Gamma_{+ }(t)-\Gamma_{- }(t)]$, can then be directly expressed as a function of the braiding phase:

\begin{eqnarray}
\Gamma_{\pm }(t) & =& C \; \Re\big[\int_{-\infty}^{t} dt' e^{\pm  i \theta \times N(t',t) } \mathcal{G}_{\delta}^2(t-t')  \big] \label{eq1}, \\
I_{T }(t) & =& -2 e^* C \; \int_{-\infty}^{t} dt' \sin{\big[\theta \times   N(t',t)\big]}    \Im \big[\mathcal{G}_{\delta}^2(t-t') \big] \label{eq2}
\end{eqnarray}
where $C$ is a constant. As discussed above, the accumulated braiding phase  $\theta \times N(t',t)$ in a time window $t-t'$ is set by the number of emitted anyons $N(t',t)$ crossing the QPC between times $t'$ and $t$:  $N(t',t)=\int_{t'}^{t} I(\tau)/e^* d\tau$, where $I(t)$ is the electrical current generated by the anyon source connected at input 1. As seen on Eq.(\ref{eq1}), $\mathcal{G}_{\delta}(t-t')$ sets the characteristic time window $t-t'$ allowing for braiding mechanisms to take place.  This can be understood from the braiding picture of Fig.1b: the time uncertainty $t-t'$  for the creation of the particle-hole excitation at the QPC is typically set by the equilibrium correlation time at the edge of the FQHC $\tau_{\delta}$. The characteristic timescale on which braiding processes may take place thus increases when decreasing $T_{\text{el}}$ or decreasing the scaling dimension $\delta$.

\section*{Tunneling dynamics and Hong-Ou-Mandel (HOM) experiment}

Eq.(\ref{eq1}) shows that the tunneling dynamics are governed by two timescales. The first one is the characteristic time for the evolution of the time-dependent braiding phase  $\theta \times N(t',t)$.  It is set by the temporal width of the anyon pulses $W$. The second one is the characteristic timescale $\tau_\delta$ of the edge correlation function $\mathcal{G}_\delta$. To simplify the respective roles of these two timescales, we consider the case of very narrow pulses, $W \ll \tau_\delta$. Two different situations may occur, depending on the number of anyons in each pulse $N$. Let us first consider the simpler case where braiding is trivial, $\theta \times N = 2\pi$. Fig.2a represents the evolution of the time-dependent braiding phase $\theta \times N(t,t')$ when varying $t$ at fixed $t'<t_0$. At time $t_0$ where the pulse crosses the QPC, the braiding phase jumps from $0$ to the value $\theta \times N =2 \pi$ in the short timescale $W$ of the pulse width. As a result, the factor $\sin{\big[\theta \times N(t',t)\big]}$ in the expression of $I_{T }(t)$ in Eq.(\ref{eq2}) jumps from $0$ back to $0$ on time $W$ (see Fig.2a). This shows that when braiding is trivial, the characteristic timescale for tunneling is set by the short temporal width of the emitted pulses. This contrasts with the situation where braiding mechanisms are non-trivial: $\theta \times N < 2\pi$. In this case, the braiding phase jumps from $0$ to the value $\theta \times N < 2\pi$ when the anyon pulse crosses the QPC (see Fig.2a in the case $N=1$), and $\sin{\big[\theta \times  N(t',t)\big]}$ varies from $0$ to the constant, non-zero value $\sin{\big(\theta \times  N \big)}$ on the short time $W$. The current then falls back to $0$ on the longer timescale $\tau_\delta$ associated with the decay of the correlation function $\mathcal{G}_{\delta}(t-t')$. The characteristic timescale of the tunneling current is thus set by the long time $\tau_\delta$ on which braiding processes may take place. This is a consequence of the fundamental property of anyons: they keep a memory of the exchange processes taking place in tunneling at a QPC. This memory, that is encoded as a braiding induced phase shift, is erased on the characteristic timescale $\tau_{\delta}$ for the decay of correlations at the edge of the fractional fluid. This implies that using triggered sources of anyons, it is possible to provide direct evidence of anyon braiding in the time-domain by measuring the slow decay of $I_T(t)$ on the timescale $\tau_\delta$.

\begin{figure}
\includegraphics[width=1
\columnwidth,keepaspectratio]{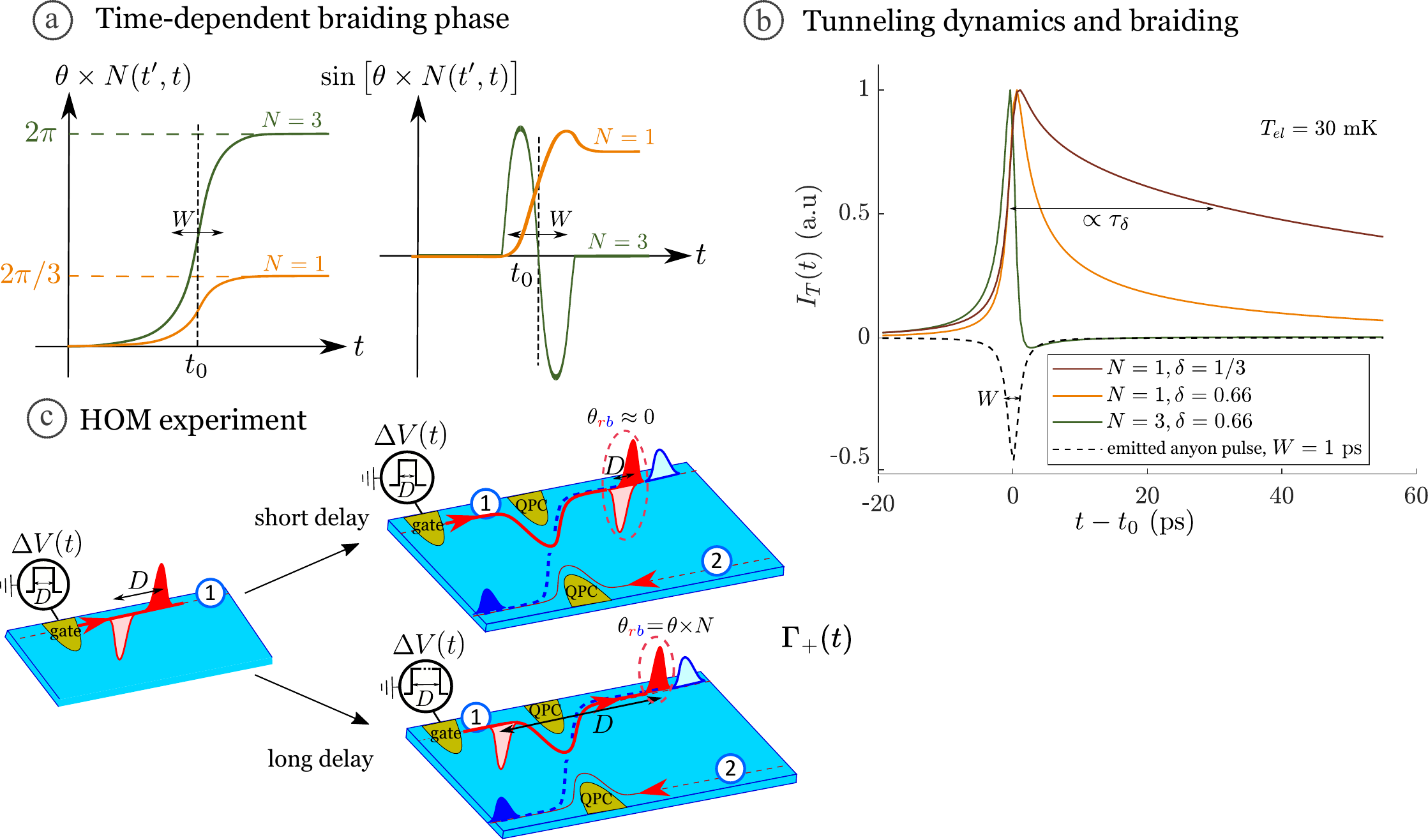}
\caption{  \small{\textbf{Time-dependent braiding phase and principle of the HOM experiment.} (\textbf{a}) Left: sketch of the time-dependent braiding phase $\theta \times N(t',t)$, where $N(t',t)$ is the number of anyons crossing the QPC between times $t'$ and $t$. It is plotted as a function of $t$ with fixed $t'<t_0$. We  compare pulses containing $N=1$ (orange trace) and $N=3$ (green trace) anyons, with $\theta = 2\pi/3$. In both cases, the braiding phase jumps from $0$ to $\theta \times N$ at the time $t=t_0$ where the anyon pulse crosses the QPC and on the timescale $W$ (width of the anyon pulse). Right: sketch of $\sin{\big[\theta \times N(t',t) \big]}$ as a function of $t$. For $N=1$, it jumps to the constant value $\sin{\big(2\pi/3 \big)}$ and stays constant afterwards, illustrating the memory left in the system for non-trivial braiding. In contrast, $\sin{\big[\theta \times N(t',t) \big]}$ varies from $0$ back to $\sin{\big(2\pi \big)}=0$ on time $W$ for $N=3$ (trivial braiding).  (\textbf{b})  Simulations of $I_T(t)$ using Eq. (\ref{eq2}). For $N=3$ (trivial braiding, green line), the current decays on the short timescale $W$, the temporal width of the emitted pulses (dashed black line). For $N=1$, the decay takes place on the longer time $\tau_{\delta}$ due to the non-trivial braiding and depends on the value of $\delta$ (1/3 for the brown line and 0.66 for the orange line).  (\textbf{c}) Principle of the HOM experiment. Two anyon pulses of opposite signs and separated by a tunable time delay $D$ are emitted at input 1 of the  QPC by applying a time-dependent square voltage to the gate.  For short time delays $D$ (upper panel), the two pulses can be considered as a single entity with zero braiding phase with the tunneling topological anyons and $\Gamma_{\pm}(t)$ are equal to their equilibrium value. For long time delays $D$ (lower panel), the two pulses give rise to two independent tunneling processes with opposite values of the braiding phase for the particle-like and hole-like pulses. This is the regime where $\Gamma_{\pm}(t)$ deviate the most from their equilibrium value. The transition between the regimes of short and long delays occurs on the characteristic timescale for $\Gamma_{\pm}(t)$ to decay down to zero: $W$ when braiding is trivial ($\theta \times N=2\pi$), $\tau_{\delta}$ when braiding is non-trivial ($\theta \times N \neq 2\pi$ ).}  }
\end{figure}

Fig.2b represents the numerically computed time-evolution of $I_T(t)$ from Eq.(\ref{eq2}) using $\theta=2\pi/3$ (the expected braiding phase at $\nu=1/3$) and for a pulse containing $N=1$ anyon ($\theta \times N=2\pi/3$). One can clearly see that the decay of $I_T(t)$ is slowed down compared to the short time $W$. It also directly depends on the value of $\delta$ as illustrated by the difference between $\delta=1/3$ (dashed orange curve) and $\delta=2/3$ (orange curve). This contrasts with the case of pulses containing $N=3$ anyons (or equivalently one electron at filling factor $\nu=1/3$) represented by the green curve on Fig.2b. In this case, braiding is trivial ($\theta \times N=2\pi$ for $\theta =2\pi/3$) and the tunneling current decays on the short timescale $W$.

Time-domain measurements of the tunneling current using fast triggered anyon sources can thus probe both the anyon braiding phase and the scaling dimension. However, measurements of the time-dependent electrical current with a time resolution of the order of $10$ ps are extremely difficult. As proposed in Ref.\cite{Jonckheere2023}, by emitting a second anyon pulse of opposite charge $-N$ at the  input of the QPC after a tunable time-delay $D$ (with a 15 picoseconds resolution), we can recover the dynamics of $\Gamma_{\pm }(t)$ from DC measurements of the time-averaged tunneling rates as a function of $D$, $\overline{\Gamma_{\pm }}(D)$. The role of the second anyon pulse is to cancel the braiding phase accumulated by the passing of the first anyon pulse (see Fig.2c). Indeed, an anyon and an antianyon fused together have zero braiding phase with a topological anyon tunneling at the QPC. However, this cancellation only occurs in the limit of short time delays (see Fig.2c, upper panel), where $D$ is much shorter than the characteristic time for $\Gamma_{\pm }(t)$ to decay down to zero: $\tau_{\delta}$ for $\theta \times N < 2\pi$ and $W$ for $\theta \times N = 2\pi$. In this limit,  the tunneling rates take their equilibrium value.  $D\gg \tau_{\delta}$ (when braiding is non-trivial) or $D\gg W$ (when braiding is trivial)  corresponds to the opposite limit of long time delays (see Fig.2c, lower panel). In this case, the two anyon pulses give rise to two independent tunneling events governed by the braiding of each pulse with the topological anyons tunneling at the QPC. This corresponds to the maximum deviation of $\overline{\Gamma_{\pm }}(D)$ from their equilibrium value. This experiment shares strong similarities with Hong-Ou-Mandel (HOM) experiments performed in optics \cite{Hong1987} or electronics \cite{Bocquillon2013} where two photons or two electrons are emitted towards a beam-splitter with a tunable time-delay between the two particles. In these two cases, HOM experiments were used to measure the temporal width $W$ of the emitted photon or electron pulses. This corresponds to the case where, for fermions and bosons, braiding is trivial. In the anyonic case, a new timescale $\tau_\delta$ appears in the dynamics of anyon transfer due to their non-trivial braiding properties.

The two experimentally accessible quantities related to  $\overline{\Gamma_{\pm }}(D)$ are the tunneling current $I_{T}(D) =e^*  (\overline{\Gamma_{+ }}-\overline{\Gamma_{- }})$ and its fluctuations $S_{T}(D) =2(e^*)^2 (\overline{\Gamma_{+ }}+\overline{\Gamma_{- }})$. Measuring both quantities simultaneously provides a stringent test of the role of braiding on the tunneling dynamics as $I_T$ and $S_T$ probe complementary aspects of the edge correlations: $I_{T}$ probes the imaginary part of $\mathcal{G}_{\delta}(t-t')$ whereas $S_{T}$ probes the real part. However, the particle-hole symmetry of the emitted anyon pulses implies that $I_{T}(D)$ vanishes for all time-delays (as $\overline{\Gamma_{+ }}=\overline{\Gamma_{- }}$). It is thus necessary to induce a small asymmetry between the two tunneling rates by applying a small DC voltage $V$ at input 2 of QPC to restore a non-zero tunneling current. The small additional stream of tunneling anyons generated by the DC voltage is then used as a non-invasive probe of the tunneling rates $\overline{\Gamma_{+/- }}$ and of their dependence on the delay $D$ via the measurement of the differential conductance $G_T(D)= I_{T}(D,V)/V$. In this work, we investigate both the excess noise $\Delta S_{T}(D)$ and the excess conductance $\Delta G_T(D)$ compared to equilibrium (anyon sources switched off, or equivalently $D=0$). Using Eq.(\ref{eq1}) for the computation of both quantities, their dependence on the time-delay $D$ allows us to probe two complementary aspects of the tunneling dynamics:
\begin{eqnarray}
\Delta S_T(D) & =& 4 C (e^*)^2 f \int_{0}^{1/f} d\bar{t} \int_{0}^{+\infty} d\tau \big[\cos{(\theta \times \Delta N(D) )} -1 \big] \Re \big[  \mathcal{G}_{\delta}^2(\tau)\big]   \label{eq3}\\
\Delta G_T(D) & = & -2 C \frac{(e^*)^2}{\hbar} f \int_{0}^{1/f} d\bar{t} \int_{0}^{+\infty} d\tau \big[\cos{(\theta \times \Delta N(D) )} -1 \big] \Im \big[  \tau \mathcal{G}_{\delta}^2(\tau)\big], \label{eq4}
\end{eqnarray}
where $\theta \times \Delta N(D)= \theta \times \int_{\bar{t}-\tau/2}^{\bar{t}+\tau/2}\frac{I(t')-I(t' -D)}{ e^* }dt'$ is the time-dependent braiding phase resulting from the two anyon pulses, and $f$ is the pulse emission frequency.

\section*{Sample}

\begin{figure}
\includegraphics[width=0.6
\columnwidth,keepaspectratio]{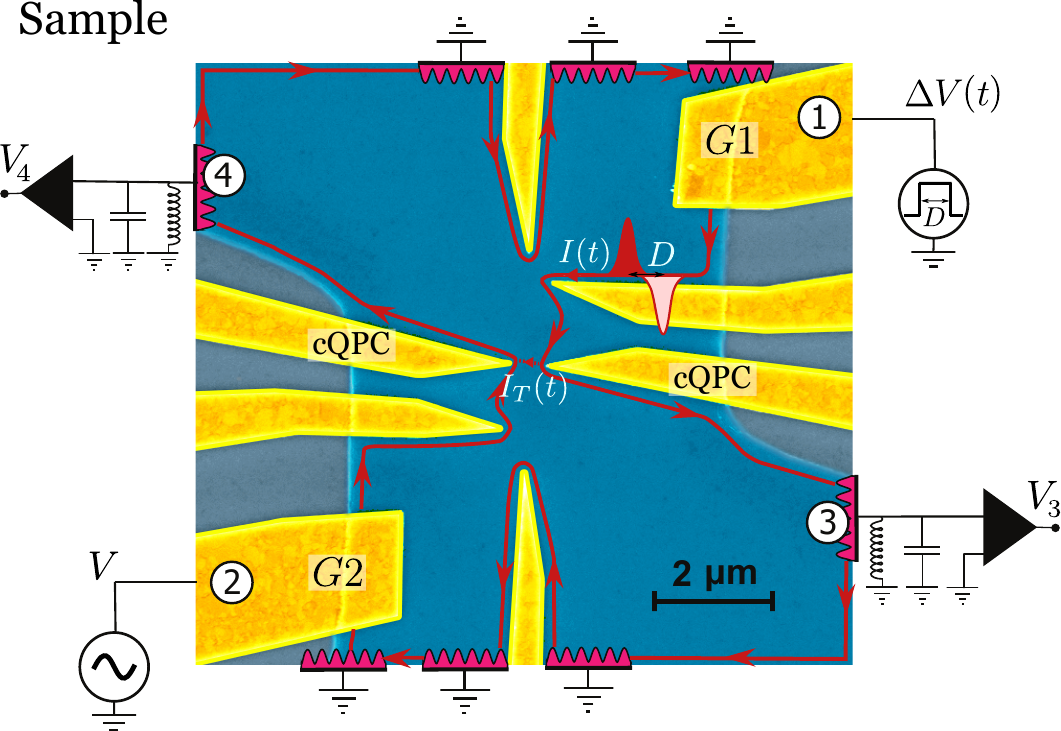}
\caption{\small{ \textbf{Sample.}  The electron gas is represented in blue, the edge channels in red and the metallic gates in gold. By applying a square voltage at a frequency $f=0.65$ GHz on gate G1,  an anyon pulse (in red) followed by a pulse of opposite charge (in light red with a red edge) are emitted at input 1 of cQPC. The time delay $D$ can be tuned by varying the width of the square excitation. The differential conductance $G_T$ is measured by applying a small AC voltage (at a frequency of 1 MHz) on gate G2 and by measuring the resulting tunneling current $I_T$ on output 3 of cQPC. The noise $S_{4}$ is measured at output 4 of cQPC using homemade cryogenic amplifiers.} }
\end{figure}

The sample, shown on Fig.3, is a two-dimensional electron gas (GaAs/AlGaAs) of density $n_s=1.1\times10^{15}$ m$^{-2}$ and mobility $\mu=1.4\times 10^6$ cm$^2$.V$^{-1}$s$^{-1}$, set at filling factor $\nu=1/3$ by applying a perpendicular magnetic field $B=13.6$ T \cite{Supp}. By applying a step voltage to the metallic gate G1, this step generates a current pulse propagating along edge channel 1 \cite{Misiorny2018}. Its charge can be tuned by the amplitude $V_{\text{exc}}$ of the excitation drive and its shape can be approximated by a square \cite{Supp} of width imposed by the transit time below the gate, $W \approx 100$ ps (with a gate length of approximately $1.7$ $\mu$m and a charge velocity of $2\times 10^4 $ m.s$^{-1}$ that we deduce from the measured pulse width). The second anyon pulse is generated by applying a second step voltage of opposite sign after a time-delay $D$ on gate G1. The emission of both pulses is then repeated at a frequency $f=0.65$ GHz, with a period $1/f \approx 1.5$ ns chosen to be much larger than both $W$ and $\tau_\delta$. The two pulses then propagate towards cQPC for the investigation of the tunneling dynamics. The current noise $S_{4}$ is measured on output 4 of cQPC. Note that $\Delta S_{4}$ differs from  $\Delta S_{T}$ by the additional contribution of the thermal noise transmitted through the QPC: $\Delta S_{4}=\Delta S_{T}-4 k_B T_{\text{el}} \Delta G_T$. This difference becomes relevant if $G_T$ varies when anyons are emitted, which is the case in the experiment. $G_T$ is measured by applying a small AC voltage at low frequency (1 MHz) on gate G2 and measuring the tunneling current through the QPC at output 3. We define the two normalized HOM experimental signals $s(D)=\frac{\Delta S_{4}(D)}{\Delta S_{4} [D\approx 1/(2f) ]}$ and $g(D)=\frac{\Delta G_T(D)}{ \Delta G_T [D \approx 1/(2f)]}$. By construction, $s$ and $g$ vary from $0$ in the short delay limit ($D=0$), to $1$ in the long delay limit ($D\approx 1/(2f)$), allowing to focus on the width of $s(D)$ and $g(D)$, which contains the information on the two characteristic timescales, $\tau_\delta$ and $W$.

\section*{Results}

We start by calibrating the charge carried by each pulse by measuring the non-normalized output noise $\Delta S_{T}[D \approx 1/(2f)]$ when the two pulses of charge $N e^*$ are maximally separated ($D \approx 1/(2f)$). Fig.4a and Fig.4b represent $\Delta S_{T}[D \approx 1/(2f)]$ and its derivative $\frac{d \Delta S_{T}[D \approx 1/(2f)]}{d V_{exc}}$ as a function of the voltage drive amplitude $V_{\text{exc}}$. $N=q/e^*$ is linearly related to $V_{\text{exc}}$ by an unknown lever arm $\alpha$,  $V_{\text{exc}}= \alpha N$. The value of $\alpha$  is adjusted for the best agreement between the data and the theory prediction of Eq.(\ref{eq3}) (see \cite{Supp} for details on the comparison with theory). As can be seen on Fig.4a, a good agreement can be obtained up to $N \approx 3$ using $\alpha = 180$ mV. In particular, it reproduces the maximum value of  $\frac{d \Delta S_{T}[D \approx 1/(2f)]}{d V_{exc}}$ observed for $N \approx 1.5$ on Fig.4b. The uncertainty on the position of the maximum is used for an estimation of the error bar on the calibration, $\alpha = 180 \pm 30$ mV. Above $N \approx 3$, discrepancies between data and model can be seen on Fig.4a and 4b, which can probably be attributed to heating effects caused by the AC excitation.

\begin{figure}
\includegraphics[width=1
\columnwidth,keepaspectratio]{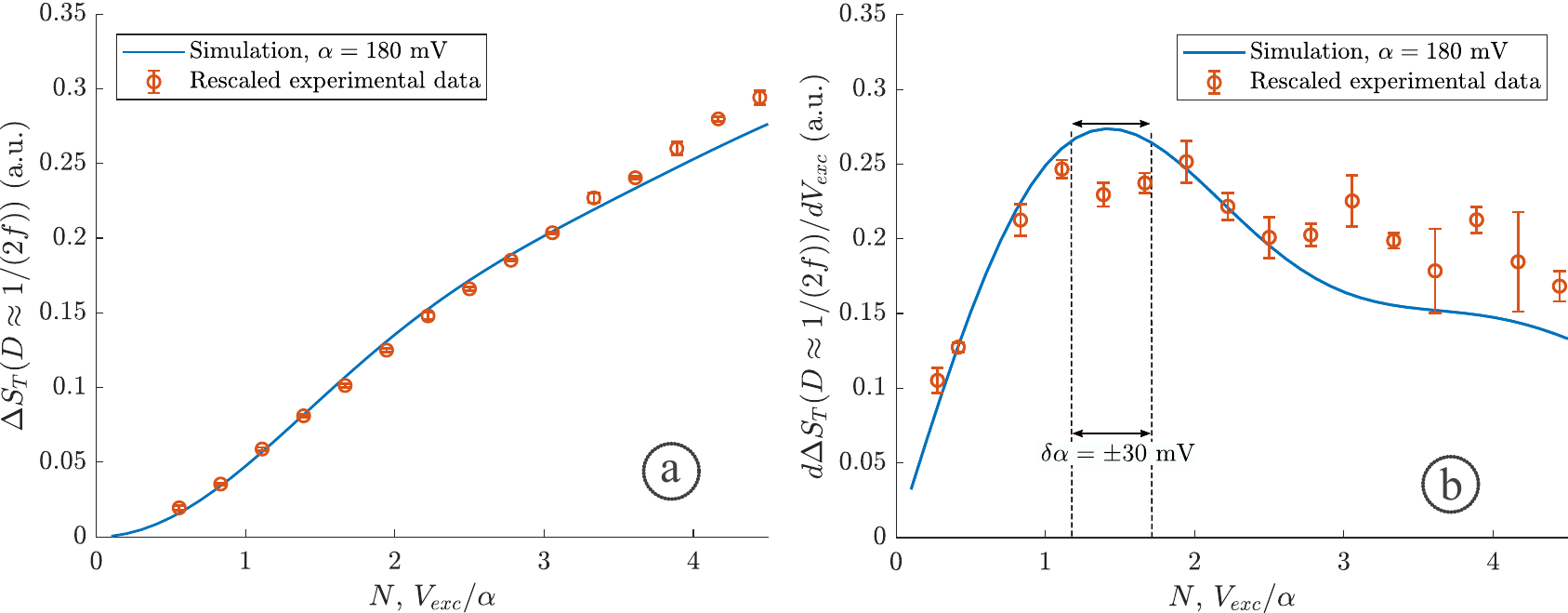}
\caption{ \small{\textbf{Calibration of the dimensionless charge $N$.} (\textbf{a}) Measurement of the output noise $\Delta S_{T}$ for perfect separation between the pulses, $D= 1/(2f)$, as a function of the amplitude $V_{\text{exc}}/\alpha = N$ of the step voltage generated by our arbitrary wave generator at room temperature. The blue trace is the numerical calculation $\Delta S_{T}(N)$ evaluated using Eq.(\ref{eq3}) with the following parameters: $T_{\text{el}}= 45$ mK, $W=123$ ps and $\delta=0.66$. The lever arm $\alpha= 180 \pm 30$ mV is adjusted for the best agreement with the data for $N \leq 3$. (\textbf{b}) $d \Delta S_{T}/d V_{\text{exc}}$ as a function of $V_{\text{exc}}/\alpha$. The blue trace is the numerical calculation using Eq.(\ref{eq3}) with the same parameters as above. The maximum of $d \Delta S_{T}/d V_{\text{exc}}$ at $N \approx 1.5$ is used for the calibration of $\alpha$.}}
\end{figure}

\begin{figure}
\includegraphics[width=1
\columnwidth,keepaspectratio]{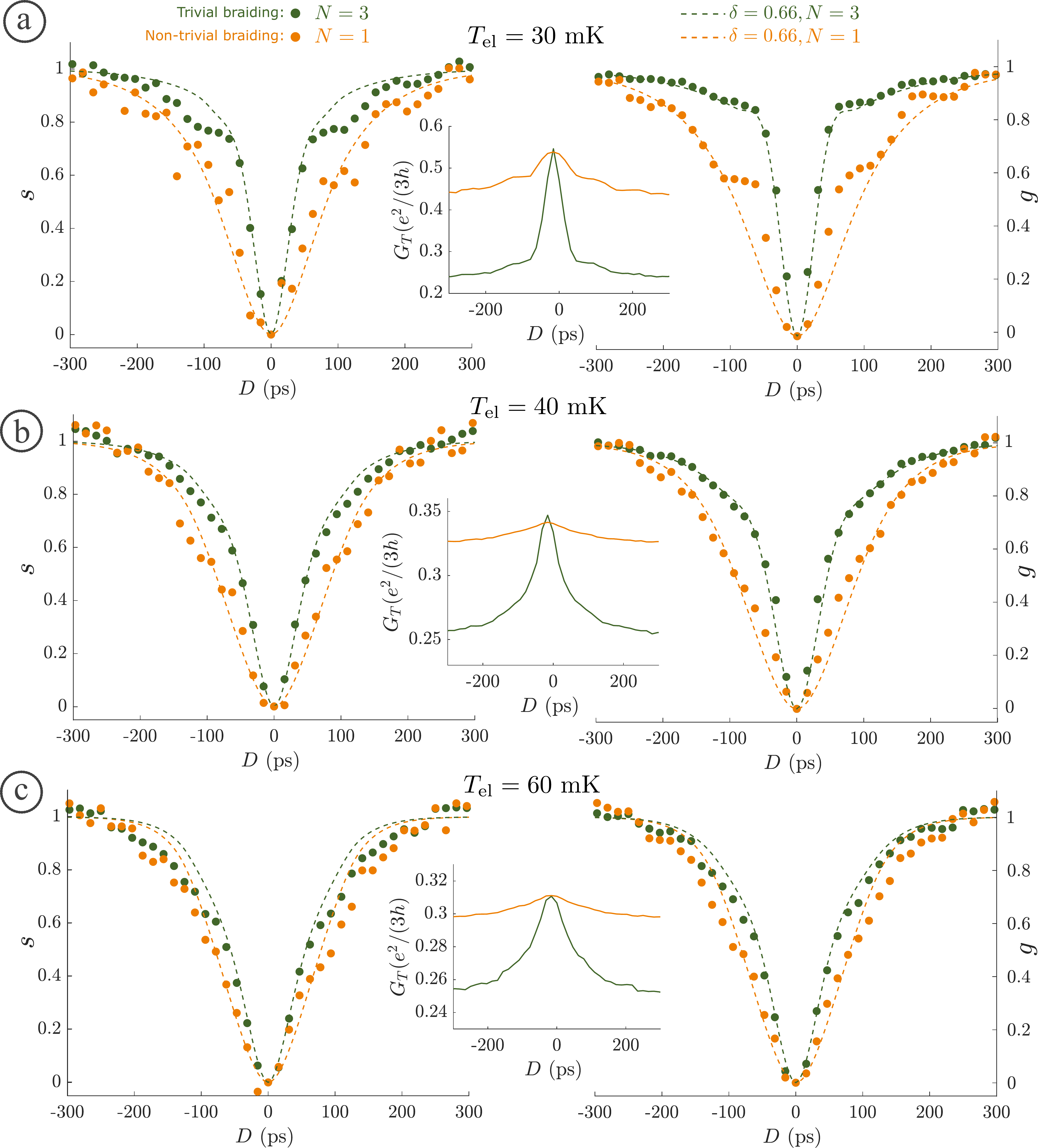}
\caption{ \small{\textbf{Braiding of anyons in the time-domain.} Measurements of $s(D)$ (left panels) and $g(D)$ (right panels) for pulses of charge $N=1$ (orange points) and $N=3$ (green points) at different temperatures. A clear widening of the HOM dip can be observed between $N=1$ and $N=3$ at the lowest temperature, $T_{\text{el}} = 30$ mK (panel (\textbf{a})). It demonstrates that non-trivial braiding increases the characteristic tunneling timescale. In other words, when braiding is present, tunneling events keep occurring long after the anyon pulse has exited the QPC. The width difference between $N=1$ and $N=3$ progressively decreases when increasing the temperature to $T_{\text{el}} = 40$ mK on panel (\textbf{b}) and $60$ mK on panel (\textbf{c}). It reflects that the characteristic timescale of braiding effects $\tau_\delta$ varies as $1/T_{\text{el}}$. Measurements are well reproduced by numerical simulations (green and orange dashed lines) of Eqs.(\ref{eq3}) and (\ref{eq4}) with the following parameters: $W=0.065/f=100$ ps, $\delta=0.66$ for panel (\textbf{a}) and  $W=0.083/f=128$ ps, $\delta=0.66$ for panels (\textbf{b}) and (\textbf{c}). The central insets represent the dimensionless differential conductance $G_T/(e^2/3h)$ as a function of $D$.} }
\end{figure}

We now turn to the measurements of the HOM traces $s$ and $g$ in the two different situations where braiding is non-trivial ($\theta \times N < 2\pi$) and where braiding is trivial ($\theta \times N = 2\pi$). For the first case we choose $N=1$, where a single anyon excitation is carried by each pulse, and that reproduces the conditions of collider experiments. For the second case we choose $N=3$, which is expected to cancel braiding effects for $\theta =2\pi/3$ at $\nu=1/3$. This can also be understood as $N=3$ corresponding to a single electron per pulse, electrons having a $2\pi$ braiding phase with anyon excitations at $\nu=1/3$. Fig.5 presents the measurements of $s(D)$ (left panel) and $g(D)$ (right panel) for three different temperatures, $T_{\text{el}}=30$ mK (in Fig.5a), $T_{\text{el}}=40$ mK (in Fig.5b) and $T_{\text{el}}=60$ mK (in Fig.5c). The green points correspond to $N=1$ anyon per pulse and the orange points to  $N=3$ anyons. All traces exhibit the common features of the HOM effect, with a dip in $s(D)$ and $g(D)$ located around $D=0$. At $T_{\text{el}}=30$ mK, a clear difference can be seen between the $N = 1$ and $N=3$ cases. The HOM dip in the $N = 3$ case is much more narrow compared to $N=1$, showing that the characteristic timescale for tunneling is shorter for $N=3$ (where braiding is trivial) compared to $N=1$ (where braiding is non-trivial). For a quantitative measurement of the width difference, we extract the width at half height $D_{1/2}$ of $s(D)$ for $N=1$ and $N=3$ from a Lorentzian fit. For $N=3$, the narrow HOM dip $D_{1/2}(N=3)=40 \pm 1.5$ ps reflects the temporal width $W$ of the current pulses propagating towards the QPC. For $N=1$, the dip increases to $D_{1/2}(N=1)=79 \pm 4$ ps, providing a direct signature, in the time-domain, of the role of braiding effects on anyon tunneling. We observe that, at the lowest temperature reached in our experiment, braiding effects increase  the characteristic tunneling timescale by a factor of 2. This effect can be seen on both quantities $s$ and $g$, with different measured shapes of the HOM dips $s(D)$ and $g(D)$. These differences reflect the fact that  $s$ and $g$ probe different aspects of the edge dynamics ($\Re[\mathcal{G}_{\delta}^2(\tau)]$ for $s$ and $\Im[\tau \mathcal{G}_{\delta}^2(\tau)]$ for $g$, see Eqs.(\ref{eq3}) and (\ref{eq4})). By increasing the temperature, the differences between $N=1$ and $N=3$ tend to be suppressed and become very faint at $T_{\text{el}}=60$ mK. This decay is expected as $\tau_{\delta}=\hbar/(\pi k_B T_{\text{el}} \delta)$ decreases when increasing the temperature, so that the widening of $I_T(t)$ due to braiding effects vanishes. For a quantitative test of the braiding scenario, all HOM traces $s(D)$ and $g(D)$ for $N=1$ and $N=3$ as well as for different temperatures are compared with numerical evaluations of Eqs.(\ref{eq3}) and (\ref{eq4}) (dashed lines, see \cite{Supp} for details of the comparison with theory). We take $\mathcal{G}_{\delta}(\tau)$ from the Luttinger model \cite{Wen1991}: $\mathcal{G}_{\delta}(\tau) \propto \frac{1}{\big[\sinh{[i \pi k_B T_{\text{el}} (\tau_0 +i \tau) /\hbar]}\big]^\delta}$, where $\tau_0$ is a short time cutoff and we take $\delta=0.66$. The only remaining unknown parameter is the width $W$ of the emitted current pulses. It is adjusted for the best agreement of $g(D)$ for $N=3$, giving $W = 100$ ps at $T_{\text{el}}=30$ mK and $W = 128$ ps at $T_{\text{el}}=40$ mK and $T_{\text{el}}=60$ mK. The discrepancy between the two values can be explained from the fact that the measurements at $T_{\text{el}}=40$ mK and $T_{\text{el}}=60$ mK were taken in the same experimental run, whereas the measurements at $T_{\text{el}}=30$ mK were obtained in a subsequent run after the removal of bias tees at cryogenic temperatures to generate sharper excitation pulses. The overall agreement is very good.  In particular, the differences between the dip widths for $s$ and $g$, as well as the suppression of braiding effects when increasing the temperature are well captured. This demonstrates that for $N=1$, braiding mechanisms slow down the tunneling dynamics compared to the $N=3$ case where braiding is trivial. Some discrepancies can be observed, which are most pronounced at $T_{\text{el}}=30$ mK. They are probably related to some additional structure on the emitted current pulses that can easily occur when sending ultrashort pulses in a quantum conductor.

\begin{figure}*
\includegraphics[width=1
\columnwidth,keepaspectratio]{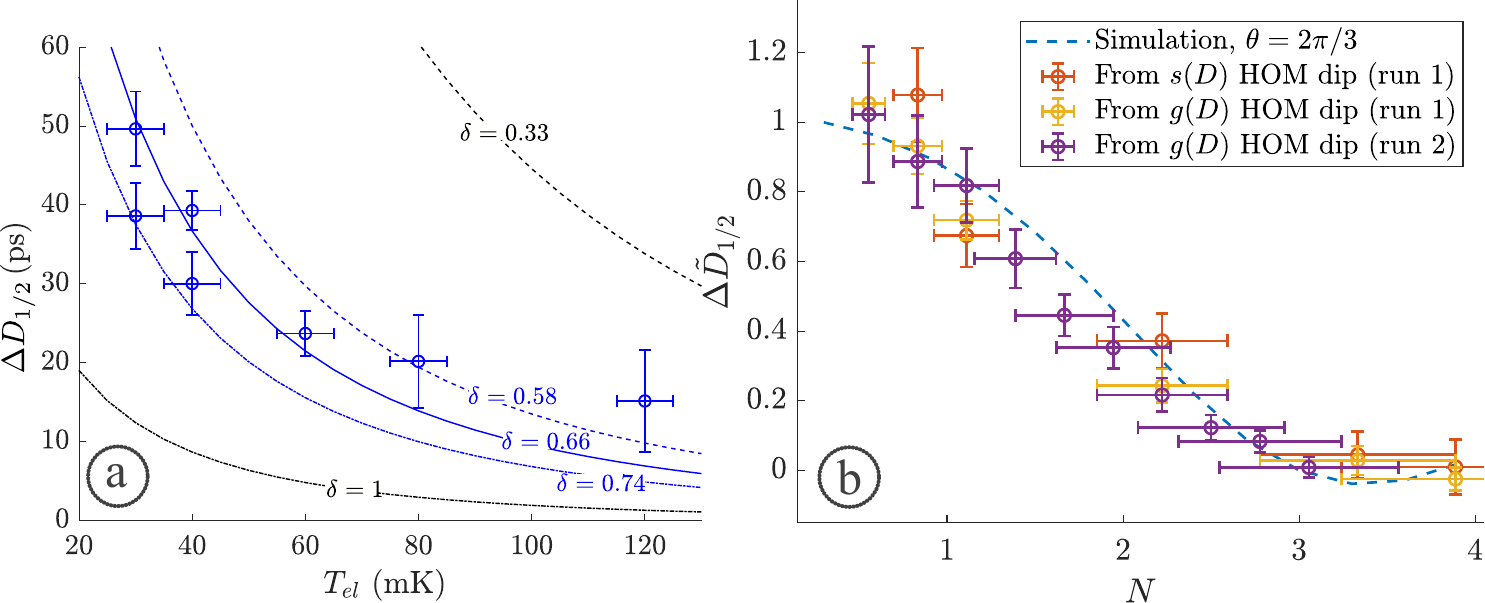}
\caption{ \small{\textbf{Extraction of the scaling dimension $\delta$ and braiding phase $\theta$.} (\textbf{a}) Excess width $\Delta D_{1/2}= D_{1/2}(N=1)-D_{1/2}(N=3)$ of the HOM dips of $s(D)$ as a function of the temperature $T_{\text{el}}$. $\Delta D_{1/2}$ decreases when increasing the temperature. This shows that the memory of braiding effects is erased on a timescale set by the temporal correlations at the edge of the quantum fluid that are inversely proportional to the temperature and to the anyon scaling dimension $\delta$.  The lines are numerical simulations with $W=100$ ps, $\delta =0.66$ (blue line), $\delta=0.58$ (blue dashed line), $\delta =0.74$ (blue dotted line), $\delta=1/3$ (black dashed line) and $\delta=1$ (black dotted line). The best agreement is obtained for  $\delta =0.66 \pm 0.08$.   (\textbf{b}) Normalized excess width $\Delta \tilde{D}_{1/2}= \frac{D_{1/2}(N)-D_{1/2}(N=3)}{D_{1/2}(N \ll 1)-D_{1/2}(N=3)}$ as a function of the anyon charge $N$. The red points are measurements of $s$ at $T_{\text{el}}=40$ mK (and $W=128$ ps), the yellow points are measurements of $g$ at $T_{\text{el}}=40$ mK (and $W=128$ ps), the violet points are measurements of $g$ at $T_{\text{el}}=30$ mK (and $W=100$ ps). The width is predicted to be minimum when braiding effects are suppressed ($\theta \times N=2\pi$) allowing for a determination of the braiding phase $\theta$.  Our measurements are in good agreement with the prediction for $\theta=2\pi/3$ (blue dashed lines). Parameters for the numerical simulation are $T_{\text{el}}=35$ mK, $W=100$ ps and $\delta=0.66$. }}
\end{figure}

For an accurate estimation of the scaling dimension $\delta$, we have measured the evolution of the width at half height $D_{1/2}(N=1)$ of the HOM dip of $s(D)$ at $N=1$  as a function of the temperature. To eliminate the contribution of the width of the emitted current pulses $W$, we plot on Fig.6a $\Delta D_{1/2}=D_{1/2}(N=1)-D_{1/2}(N=3)$ as a function of the temperature $T_{\text{el}}$. As discussed above, $\Delta D_{1/2}$ decreases when increasing the temperature and the decay rate can be directly compared to the predictions of Eq.(\ref{eq3}) for different values of $\delta$ ($\delta=0.66$ is the blue line, $\delta=0.58$ the blue dashed line and $\delta=0.74$ the blue dotted line). The best agreement with theory is obtained for $\delta = 0.66 \pm 0.08$.

In order to probe the value of the braiding phase $\theta$ at $\nu=1/3$ from our time-domain experiment, we measure the continuous evolution of the width of the HOM dip for $s$ and $g$ as a function of the dimensionless charge $N$ carried by the anyon pulses. As the widths of the HOM dips for $s$ and $g$ are different (see Fig.5), we define the normalized width $\Delta \tilde{D}_{1/2}(N)=\frac{D_{1/2}(N)-D_{1/2}(N=3)}{D_{1/2}(N \ll 1)-D_{1/2}(N=3)}$ which takes almost identical values for $s$ and $g$ (the small differences predicted by theory are much smaller than the experimental resolution). Fig.6b gathers the measurements of the normalized excess width for the HOM traces of both the noise $s(D)$ (red points) and  the differential conductance $g(D )$ (violet and yellow points). For $N\leq 1$, the normalized width is close to $1$ and then decays smoothly towards $0$ as $N$ reaches $3$. This behavior is well reproduced by the model (blue dashed line) using $\theta = 2\pi/3$. This shows that braiding effects can be suppressed by generating pulses containing $N=3$ anyons, implying $\theta \times 3 =2\pi$ and thus confirming the value $\theta=2\pi/3$ of the braiding phase at $\nu=1/3$. This demonstrates the value of time-resolved measurements for probing the properties of anyons. A better accuracy for the measurement of the braiding phase could be obtained by improving the calibration of the number of anyons per pulse $N$, and by generating anyon pulses with shorter temporal width $W$. Indeed, when varying $N$, the sharpness of the transition between the regime dominated by braiding ($\Delta \tilde{D}_{1/2}(N)=1$) and the regime where braiding is trivial ($\Delta \tilde{D}_{1/2}(N)=0$) is directly controlled by the temporal width of the anyon pulses \cite{Supp}: the sharper the pulses are, the more abruptly the cancellation of braiding effects happens.

We finally investigate how the HOM experimental signals vary when tuning the QPC gate voltage that sets the QPC backscattering probability. For this study, we focus on the measurement of $g$, which are faster to carry out (an additional measurement of $s$ for $G_T < 0.1 \frac{e^2}{3h}$ is shown in \cite{Supp}). Fig.7a presents all our measurements of $G_T(D)$ for $N=1$ and $N =3$ and for different settings of the QPC backscattering probability $P_0$ defined at zero time-delay: $P_0=G_T(D=0)/(e^2/3h)$. Interestingly, the measurements of  $G_T(D)$ for $P_0>0.1$ agree qualitatively with the predictions of the Luttinger model for the non-linear evolution of the differential conductance $G_T$ (decrease of $G_T$ when increasing the time-delay $D$). In contrast, the measurements of $G_T(D)$ for $P_0=0.06$ exhibits an anti-Luttinger dependence ($G_T$ increases when increasing $D$). This behavior, which is systematically observed in the very weak backscattering regime in I-V characteristics measurements would suggest an inconsistent value of $\delta >1$. In contrast, all our measurements of $g(D)$ for $N=1$ and $N=3$ plotted on Fig.7b collapse on the same curves, with the same increase of the width of the HOM dips for all values of $P_0$. This shows that by probing the scaling dimension directly in the time-domain, our measurement of $\delta \approx 0.66$ is robust to the variations of the QPC backscattering probability $P_0$.

\begin{figure}
\includegraphics[width=0.9
\columnwidth,keepaspectratio]{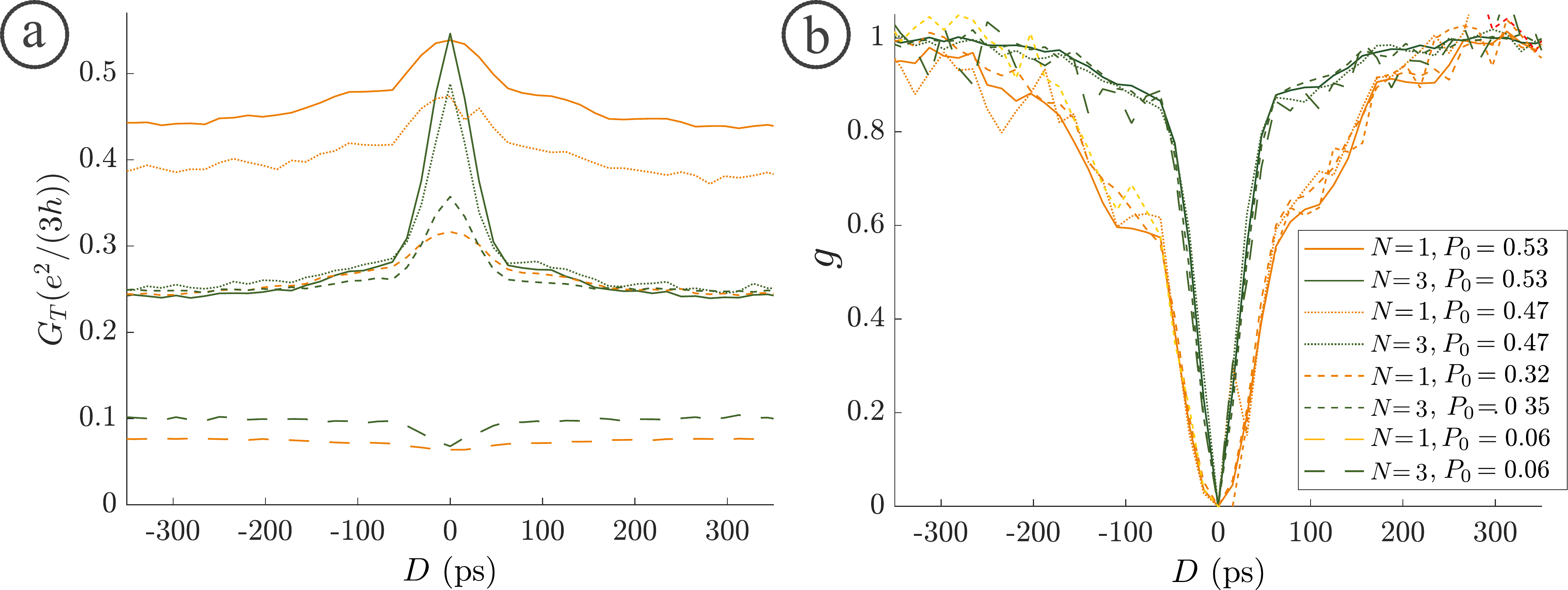}
\caption{ \small{\textbf{Evolution of the scaling dimension with the QPC backscattering probability. } (\textbf{a})  Measurements of $G_T(D)$ for $N =1$ and $N=3$ for different values of the backscattering probability. (\textbf{b})  Normalized dimensionless conductance $g(D)$ for $N =1$ and $N=3$ corresponding to the measurements of  $G_T(D)$ plotted on panel (\textbf{a}). Despite the measured differences between the different traces of  $G_T(D)$, all traces of $g(D)$ for $N=1$ and $N=3$ fall on the same curve. In particular, the difference of the width of the HOM dips between $N=1$ and $N=3$ is the same for the different traces corresponding to different settings of the QPC backscattering probability. It  shows that the scaling dimension $\delta$ is rather insensitive to the specific value of the QPC backscattering probability.} }
\end{figure}

\section*{Conclusion}

To conclude, we have investigated the scaling dimension and braiding phase of topological anyons tunneling at a QPC at filling factor $\nu=1/3$ by studying the partitioning of short pulses carrying a tunable number of anyons. By emitting two mutually time-delayed anyon pulses of opposite charge in a HOM setup, we have measured the characteristic timescale for tunneling at the QPC. This is performed by measuring the width of the HOM dips of two experimental quantities: the current fluctuations at the QPC output $s(D)$, and the differential conductance $g(D)$, as a function of the time-delay between the two pulses $D$. We have characterized the dependence of $s(D)$ and $g(D)$ on the number of anyons per pulse $N$ that controls the mutual braiding between the emitted anyon pulses and the topological anyons tunneling at the QPC.  We have compared two situations, $N=1$ where braiding is non-trivial (with a mutual braiding phase $\theta \times 1= 2\pi/3$ at $\nu=1/3$), and $N=3$ where braiding effects are expected to be suppressed at $\nu=1/3$ ($\theta \times 3= 2\pi$). At the lowest temperature $T_{\text{el}}=30$ mK, we observe a widening by a factor 2 of the HOM dip between $N=3$ and $N=1$. It shows that non-trivial braiding increases the characteristic tunneling time, being governed at the edge of the FQHC by the temporal correlations $\tau_{\delta}=\hbar/(\pi k_B T_{\text{el}} \delta)$. This provides the first experimental signature of the role of braiding on anyon dynamics, strikingly illustrating that anyons have specific physical properties, distinct from those of fermions or bosons. By increasing the temperature, we observe that the differences between $N=1$ and $N=3$ are suppressed, as expected from the $1/T_{\text{el}}$ dependence of $\tau_{\delta}$. From the temperature dependence of the width difference $\Delta D_{1/2}(T_{\text{el}})$ of the HOM dips between $N=1$ and $N=3$, we extract the scaling dimension $\delta=0.66 \pm 0.08$. By measuring the evolution of the normalized width $\Delta \tilde{D}_{1/2}(N)$ with the number of anyons per pulse $N$, we characterize the suppression of braiding effects on the dynamics of anyon tunneling when $\theta \times N$ approaches $2\pi$. Our measurements show that braiding becomes trivial for $N=3$, confirming previous measurements of $\theta = 2\pi/3$ at $\nu=1/3$ \cite{Bartolomei2020,Nakamura2020,Ruelle2023,Glidic2023,Lee2023}. Our measured value of $\delta=0.66 \pm 0.08$ contrasts with the expectation of $\delta=\nu=1/3$ for a Laughlin FQHC, showing that $\delta$ is non-universal\cite{Rosenow2002}. Our experiment raises the question of the control and engineering of the dynamical properties of the edge channels set by $\delta$: can it be modified by changing the nature of the edge, the quantum point geometry, or the nature of the conductor (GaAs versus graphene)? Regarding the characterization of anyon properties in FQHC, our time-domain experiment allows us to disentangle the effect of the scaling dimension $\delta$, that controls the characteristic timescale when braiding is non-trivial, from that of the braiding phase $\theta$, that controls the non-trivial or trivial nature of braiding mechanisms in tunneling experiments. It will thus be instrumental for the characterization of anyon braiding in more complex abelian FQH states where the fractional charge $e^*$ and the braiding phase $\theta$ are given by different quantized numbers. For example, at $\nu=2/5$, anyons have a fractional charge $e^*=e/5$ and a braiding phase $\theta = 3/5 \times 2\pi$, meaning that pulses containing $N=5/3$ anyons would be necessary to cancel braiding effects ($3/5 \times 2\pi \times 5/3 =2\pi$). This would correspond to anyon pulses containing a charge $q=N e/5=e/3$. Beyond abelian FQH states, our time-domain measurements will also offer a new approach for probing the dynamical properties of non-abelian states \cite{Willett1987,Moore1991},  providing new information and constraints on the nature of their ground state and excitations.



\section*{Acknowledgments }
\textbf{Funding:} This work has been supported by the  ERC advanced grant ``ASTEC'' (grant No. 101096610), the ANR grant ``ANY-HALL'' (ANR-21-CE30-0064),
and by the French RENATECH network.

\textbf{Author Contributions:} YJ fabricated the sample on GaAs/AlGaAs heterostructures grown by AC and UG. YJ designed and fabricated the low-frequency cryogenic amplifiers used for noise measurements. MR and EF conducted the measurements. MR, EF, EB, JMB, BP, GM and GF participated to the data analysis and the writing of the manuscript with inputs from BG, TJ, TM, JR, YJ and UG. Theory was done by BG, TJ, TM and JR.  GF and GM supervised the project. 

\textbf{Competing interests:} The authors declare no competing interests.  \textbf{Data and materials availability:} All data is available in the manuscript or the supplementary materials.

\textbf{Note added:} Coincident to the present investigation, two other works are experimentally addressing the scaling dimension of the $e/3$ fractional quantum Hall quasiparticles at $\nu=1/3$ using a different technique: the thermal to shot noise crossover of the current fluctuations at the output of a QPC driven by a DC voltage. An experiment by the team of M. Heiblum with a theoretical analysis led by K. Snizhko (N. Schiller et al. \cite{Schiller2024}) finds $\delta=1$. The team of F. Pierre and A. Anthore finds $\delta=1/3$, in agreement with expectations at $\nu=1/3$ (A. Veillon, et al. \cite{Veillon2024}). The differences observed between the various works shows the non-universality of $\delta$ that may be related to differences in the geometry of the QPCs.

\newpage
\begin{center}
  \huge{Time-domain braiding of anyons: supplementary notes}  
\end{center}

\section{ Current pulses generated by applying a voltage drive to a gate}

\subsection{Bosonic field}

We consider a single edge channel at filling factor $\nu=1/3$. The dynamics at the edge are described by a bosonic field $\phi(x)$ with the following commutation relations:
\begin{equation}
[\phi(x), \phi(x')]  =  i \pi \text{Sign}(x-x') ,
\end{equation}
where $\text{Sign}(x)$ is the sign function.
The creation operator for an anyon then reads
\begin{equation}
\Psi^\dagger (x) \propto e^{i \sqrt{\nu} ~\phi(x)} .
\label{eq:commphi}
\end{equation}
Anyon pulses are generated by applying an external drive $V_g(x,t)$ on a gate, leading to the shift of the bosonic field:
\begin{equation}
\phi(x,t)  \longrightarrow \phi(x,t) +
 \frac{ \sqrt{\nu} e}{\hbar} \int_{-\infty}^t dt' V_g (x - v (t-t'), t') ,
\end{equation}
where $v$ is the propagation speed of edge magnetoplasmon modes.
Several interesting limits can then be discussed:

\subsection{Ohmic contact or semi-infinite gate}

The application of a voltage drive $V(t)$ to a contact connected to a single edge channel can be modeled by the application of a drive $V_{g}(x,t)=V(t) \theta(-x)$ to a semi-infinite gate that would extend for $x \in ]-\infty, \; 0]$ ($\theta$ is the Heaviside step function). This leads to the following shift or displacement of the bosonic field for $x>0$:
\begin{eqnarray}
\phi_{\infty} & =&  \frac{\sqrt{\nu} e}{\hbar} \int_{-\infty }^{t} dt' V(t') \theta(v (t-t')-x)  \\
 & =&
 \frac{ \sqrt{\nu} e}{\hbar} \int_{-\infty}^{t-x/v} dt' V (t') .
 \label{eq:phidis}
\end{eqnarray}

The current at position $x>0$ and time t is then related to the bosonic field by $I(x,t)= -e v \rho(x,t)$ where $\rho(x,t) = \frac{\sqrt{\nu}}{2\pi} \partial_x \phi(x,t)$ is the particle density, and one recovers the usual expression for the conductance:
\begin{eqnarray}
I(x,t)=  \frac{\nu e^2 }{h} V(t-x/v) .
\end{eqnarray}

\subsection{Infinitely narrow gate}

For an infinitely narrow gate placed at $x=0$, the same current pulse $I(x,t)= \frac{\nu e^2 }{h} V(t-x/v)$ can be created along the bosonic edge after the gate by applying the external drive:
\begin{equation}
V_{ng}(x,t) = v \, \delta(x) \int_{-\infty}^{t} d \tau V(\tau),
\end{equation}
where the applied voltage to the narrow gate is the integral of the voltage that would be applied to an ohmic contact to generate the same current pulse.
It indeed creates the same shift of the bosonic field:
\begin{align}
\phi_{ng} &= \frac{ \sqrt{\nu} e}{\hbar} \int_{-\infty}^t dt' V_{ng} (x - v (t-t'), t') \\
&= \frac{ \sqrt{\nu} e}{\hbar} \int_{-\infty}^t dt' v \delta (x - v (t-t')) \int_{- \infty}^{t'} d\tau V(\tau)  \nonumber \\
&= \frac{ \sqrt{\nu} e}{\hbar}  \int_{- \infty}^{t - \frac{x}{v}} d\tau V(\tau) .
\label{eq:Phideltares}
\end{align}

\subsection{Finite size gate} \label{sec:fsize}

We now consider a realistic gate having a finite width $a$. The external drive
applied uniformly on the gate thus becomes
\begin{equation}
V_{a}(x,t) = v \, R(x,a) \int_{-\infty}^{t} d \tau V(\tau),
\end{equation}
where $R(x,a)= \frac{1}{a} (\theta(x+a/2)-\theta(x-a/2))$ is the rectangular window function of
width $a$.
This leads to the following shift:
\begin{align}
\phi_a=\frac{ \sqrt{\nu} e}{\hbar}  \int_{-\infty}^t dt' V_{a} (x - v (t-t'), t') =
\frac{ \sqrt{\nu} e}{\hbar} \int_{-\infty}^t dt' v R(x - v (t-t'),a) \int_{- \infty}^{t'}  d\tau V(\tau) .
\end{align}
As the window function $R$ is non-zero only for $t'$
in the interval $[t-x/v-a/(2 v),t-x/v+a/(2 v)]$, and
constant in this interval, the shift
is equal to:
\begin{equation}
\phi_a= \frac{ \sqrt{\nu} e}{\hbar}  \frac{1}{a} \int_{-a/2}^{a/2}  d\epsilon
  \int_{-\infty}^{t-\frac{x+\epsilon}{v}}  d\tau  V(\tau)   .
\end{equation}
We are interested in the regime where the travel time of the gate width
$a/v$ is much larger than the width of the pulses of $V(t)$. For simplicity,
we can then consider that $V(t) = V_{exc}  \delta(t)$ is a delta peak at $t=0$. The shift is then:
\begin{align}
\phi_a &= \frac{ \sqrt{\nu} e}{\hbar}  \frac{1}{a} \int_{-a/2}^{a/2} d\epsilon \int_{-\infty}^{t-\frac{x+\epsilon}{v}}  d\tau   V_{exc} \delta \left( \tau  \right) \nonumber \\
&= \frac{ \sqrt{\nu} e}{\hbar}  \frac{1}{a} \int_{-a/2}^{a/2}  d\epsilon \int_{-\infty}^{t-\frac{x}{v}}  d\tau   V_{exc} \delta \left( \tau - \frac{\epsilon}{v} \right) \nonumber \\
&= \frac{ \sqrt{\nu} e}{\hbar}  \int_{-\infty}^{t-\frac{x}{v}} d\tau  V_{exc} \frac{1}{a}  \int_{-a/2}^{a/2} \!\! d\epsilon  \; \delta \left( \tau - \frac{\epsilon}{v} \right) \nonumber \\
&= \frac{ \sqrt{\nu} e}{\hbar}   \int_{-\infty}^{t-\frac{x}{v}}  d\tau  V_{exc}   R\left(\tau ,\frac{a}{v} \right) ,
 \label{eq:Phiares}
\end{align}
where we used the fact that the window function of width $a$ is recovered by averaging a delta function over this width. The result of Eq.~(\ref{eq:Phiares}) is similar to Eq.~(\ref{eq:Phideltares}), obtained for an infinitely narrow gate, but with the peaks of $V(t)$ replaced by square peaks of width $a/v$ (see Fig.S1).
Eq.(\ref{eq:Phiares}) is valid as long as $a/v$ is much larger than the width of the applied $V(t)$, which corresponds to the experimental situation.

\begin{figure}[h]
    \centering
    \includegraphics[scale=0.5]{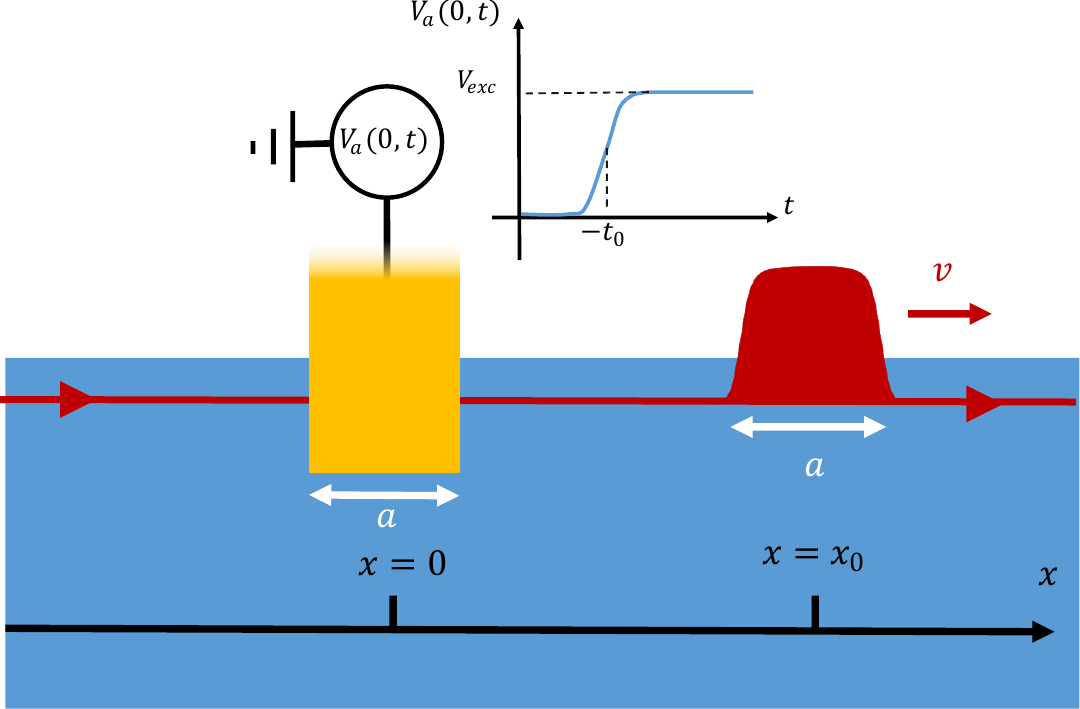}
    \legend{\small{\textbf{Fig.S1: emission of an anyon pulse at position $x_0$ using a finite size gate.} Following Eqs. (S9) and (S12), a square excitation pulse $V_a(x=0,t)$ generates a rectangle anyon pulse (or square peak) with a spatial width limited by the gate size $a$ corresponding to a temporal width $a/v$. For an excitation voltage pulse centered around time $t=-t_0$, the anyon pulse reaches position $x_0=v t_0$ at time $t=0$.   } }
    \label{fig:sweep}
\end{figure}

\section{Braiding of anyon pulses at a QPC}

We now consider the creation of an anyon current pulse of charge $q = N e^*$ and small width $W$ that reaches the position $x_0=v t_0>0$ at time $t=0$ (as represented on Fig.S1). It means that the applied voltage pulse is centered around time $t=-t_0$ (with $t_0 \gg W$) and satisfies $\int d\tau I_N (\tau)= Ne^*$, where $I_N (\tau)$ is the pulse electrical current.
The anyon creation operator $\Psi^{\dag}_N$ associated with this pulse, is the displacement operator of the bosonic field:
\begin{align}
\Psi^{\dag}_N (x_0) & = e^{\frac{i \sqrt{\nu} e}{h} \int_{-\infty}^{+\infty} d\tau V_{N}(\tau) \phi(-v \tau) } .
\label{eq:dis}
\end{align}

The shift $\phi_{N}$ of the bosonic field  is obtained from the commutator:
\begin{equation}
[\phi(x),\Psi^{\dag}_{N}(x_0)] = \phi_{N} \Psi^{\dag}_{N}(x_0)  .
\end{equation}
We use
\begin{equation}
[A,F(B)] = [A,B] \, F'(B)
\end{equation}
with $A=\phi(x)$ and
$B=\frac{i \sqrt{\nu}  e}{h} \int_{-\infty}^{+\infty} d\tau V_{N}(\tau) \phi(-v \tau)$, and which is valid as both $A$ and $B$ commute with their commutator $[A,B]$, to get:
\begin{align}
[\phi(x),\Psi^{\dag}_{N}(x_0)] & = -\frac{e \pi \sqrt{\nu} }{h} \int_{-\infty}^{+\infty} d\tau  V_{N}(\tau) \text{Sign}(x+v \tau)  \times \Psi^{\dag}_{N}(x_0) \nonumber  \\
&=  -\frac{e \sqrt{\nu} }{2 \hbar} \left\{
 \int_{-x/v}^{+\infty}  d\tau  V_{N}(\tau) - \int_{-\infty}^{-x/v}  d\tau  V_{N}(\tau) \right\} \times \Psi^{\dag}_{N}(x_0) \nonumber \\
  & = -\frac{e \sqrt{\nu} }{2 \hbar} \left\{ \int_{-\infty}^{+\infty} d\tau  V_{N}(\tau) - 2 \int_{-\infty}^{-x/v}  d\tau  V_{N}(\tau) \right\}  \times \Psi^{\dag}_{N}(x_0) .
\end{align}
The first term (equal to $-\sqrt{\nu} \pi N$ in the limit $t_0 \gg W$) is independent of $x$
and can be safely omitted. We thus get finally:
\begin{equation}
\phi_{N}  =  \frac{e \sqrt{\nu} }{\hbar} \int_{-\infty }^{-x/v} dt'V_{N}(t'),
\end{equation}
which is identical to Eq.~(\ref{eq:phidis}) for $t=0$ (the shift at an arbitrary time $t$
is simply obtained by replacing $x$ by $x-v t$, leading to Eq.~(\ref{eq:phidis})).

We now compute the commutation relation between the operator $\Psi^{\dag}_N (x_0)$ creating an anyon current pulse of charge $q = N e^*$ at position $x_0$ and the operator $\Psi^{\dag} (x_0')$ creating a topological anyon at position $x_0'$:

\begin{align}
\Psi^{\dag}_N (x_0) \Psi^{\dag} (x_0') & = \Psi^{\dag} (x_0')  \Psi^{\dag}_N (x_0)  e^{- \frac{\nu e}{h} \int_{-\infty}^{+\infty} d\tau V_{N}(\tau) [ \phi(-v \tau) , \phi(x_0')]} \nonumber \\
& = \Psi^{\dag} (x_0')  \Psi^{\dag}_N (x_0)  e^{- i \pi \frac{\nu e}{h} \int_{-\infty}^{+\infty} d\tau V_{N}(\tau) \text{Sign}(-v \tau - x_0')} \nonumber \\
& = \Psi^{\dag} (x_0')  \Psi^{\dag}_N (x_0)  e^{- i \frac{\pi}{e} \int_{-\infty}^{+\infty} d\tau I_{N}(\tau) \text{Sign}(-v \tau - x_0')} .
\label{eq:comrelPsi}
\end{align}
The current pulse $I_{N}(\tau)$ takes non-negligible values for $\tau \approx -t_0 \pm W$, so that
\begin{align}
\Psi^{\dag}_N (x_0) \Psi^{\dag} (x_0') & = \Psi^{\dag} (x_0')  \Psi^{\dag}_N (x_0)  e^{- i \frac{\pi}{e} \int_{-\infty}^{+\infty} d\tau I_{N}(\tau) \text{Sign}(x_0 - x_0')} \nonumber \\
& = \Psi^{\dag} (x_0')  \Psi^{\dag}_N (x_0)  e^{- i N \pi \nu \text{Sign}(x_0 - x_0')}  ,
\label{eq:PsiNPsi}
\end{align}
where we have used that $\int_{-\infty}^{+\infty} d\tau  I_{N}(\tau) = N \nu e$.

The result of Eq.~\eqref{eq:PsiNPsi} shows that an anyon pulse of charge $N e^*$ and a topological anyon have mutual statistics $\varphi_{N} = N \pi \nu$. The braiding phase of a topological anyon with an anyon pulse of charge $N e^*$ is thus $2\varphi_{N}=N \theta$, with $\theta = 2\pi /3$ at $\nu=1/3$ as discussed in the manuscript.

Note that taking the limit of very short pulses $W \rightarrow 0$ and  $I_{N}(\tau)= N e^* \delta (\tau+ t_0)$, we obtain:
\begin{equation}
\Psi^{\dag}_N (x_0)  = e^{i N \sqrt{\nu}  \int_{-\infty}^{+\infty} d\tau  \delta (\tau+ t_0) \phi(-v \tau) } = e^{i N \sqrt{\nu}  \phi(x_0) } ,
\end{equation}
which for $N=1$ is exactly the creation operator for a topological anyon of charge $e^*=e/3$ at $\nu=1/3$.

\section{Disentangling the statistics from other effects}

The Laughlin states are characterized by a rather simple topological order, where the effective charge and the exchange phase of topological anyons are both directly given by the filling factor $\nu$. It may thus become difficult to keep track and isolate the effect of statistics from other contributions.

Borrowing from the standard theory used to describe more complex topological orders\cite{Wen1995,Shtanko2014}, such as composite edge states, we here reformulate the results of the previous section in order to better emphasize the role of statistics.

Starting from a bosonic theory of the edge, with standard commutation relations as described by Eq.~\eqref{eq:commphi}, one may introduce the charge density as $\rho = \frac{Q}{2 \pi} \partial_x \phi$. There, $Q$ encodes how the bosonic field couples to the electromagnetic field. The conductance of the edge state is then given by $G = Q^2 e^2/h$. The conductance is related to the bulk filling fraction which, in the present case of a single edge, trivially leads back to the expected value $Q = \sqrt{\nu}$.

The edge state hosts quasiparticles whose creation operators take the general form
\begin{equation}
\Psi^\dag (x) \propto e^{i l \phi (x)} .
\end{equation}
Working out the commutation relation of two such topological anyons, one readily obtains
\begin{equation}
\Psi^\dag (x) \Psi^\dag (x') = \Psi^\dag (x') \Psi^\dag (x) e^{- i \pi l^2 \text{Sign}(x-x')} ,
\end{equation}
so that the braiding phase is given by $\theta = 2 \pi l^2$. In the single edge case, one has simply $l=\sqrt{\nu}$, which is the same
value as $Q$. Our goal here is precisely to distinguish between the contribution related to the conductance
(given by $Q$) and the contribution related to the statistics (given by $l$).

The commutation relation between this topological anyon creation operator and the one corresponding to the total charge along the edge, $\int dx \rho (x)$ leads to the expression for the effective charge of an anyon, namely
\begin{equation}
e^* = e l Q .
\end{equation}

In the presence of an external voltage drive $V_N(t)$, the bosonic field gets shifted as
\begin{equation}
\phi(x)  \longrightarrow \phi(x) + \frac{ Q e}{\hbar} \int_{-\infty}^{-x/v} dt' V_N (t') ,
\end{equation}
which can be captured by the action of a displacement operator, thus yielding the creation operator of an anyon current pulse
\begin{align}
\Psi^{\dag}_N (x_0) & = e^{\frac{i Q e}{h} \int_{-\infty}^{+\infty} d\tau V_{N}(\tau) \phi(-v \tau) } .
\end{align}
One readily sees that such an operator only involves the electromagnetic coupling $Q$ as expected.

In turn, working out the commutation relation between the creation operator associated with an anyon current pulse (of charge $N e^*$) and a topological anyon, one obtains,
 \begin{align}
\Psi^{\dag}_N (x_0) \Psi^{\dag} (x_0') & = \Psi^{\dag} (x_0')  \Psi^{\dag}_N (x_0)  e^{- \frac{Q l e}{h} \int_{-\infty}^{+\infty} d\tau V_{N}(\tau) [ \phi(-v \tau) , \phi(x_0')]} \nonumber \\
& = \Psi^{\dag} (x_0')  \Psi^{\dag}_N (x_0)  e^{- i \pi \frac{Q l e}{h} \int_{-\infty}^{+\infty} d\tau V_{N}(\tau) \text{Sign}(-v \tau - x_0')} \nonumber \\
& = \Psi^{\dag} (x_0')  \Psi^{\dag}_N (x_0)  e^{- i \pi \frac{Q l e}{h G} \int_{-\infty}^{+\infty} d\tau I_{N}(\tau) \text{Sign}(x_0 - x_0')} ,
\end{align}
 where we used that $I_{N}(\tau)$ takes non-negligible values for $\tau \approx -t_0 \pm W$.

 Substituting back the expression for the conductance $G$ along the edge, and using that $\int_{-\infty}^{+\infty} d\tau  I_{N}(-\tau) = N e^* = N l Q e$, one finally gets
\begin{align}
\Psi^{\dag}_N (x_0) \Psi^{\dag} (x_0') & = \Psi^{\dag} (x_0')  \Psi^{\dag}_N (x_0)  e^{- i N \pi l^2 \text{Sign}(x_0 - x_0')}  \label{eq:commutation},
\end{align}
 which then implies that the braiding phase between an anyon pulse of charge $N e^*$ and a topological anyon is $N \theta = N 2 \pi l^2$ (which reduces to $N \times  2 \pi \nu$ for a single edge). The braiding phase is thus directly related to the parameter $l$,
  which contains the statistical properties of the topological quasiparticles.
 This shows that measuring the braiding phase between
 an anyon pulse and a topological anyon gives access to true statistical properties
 of the topological excitations of the system.

\section{Computing the time-dependent tunneling rates}

The presence of a QPC (at position $x=0$) in the weak backscattering regime, enables the tunneling of single anyons between the two edges (labeled hereafter 1 and 2). This process is described microscopically by a tunneling Hamiltonian of the form
\begin{equation}
\hat{H}_T = \xi \left[ A_+ + A_- \right]
\end{equation}
where the tunneling operators satisfy $A_+ = A_-^\dagger$ and one has $A_-  = \Psi_1^\dagger (x=0) \Psi_2 (x=0) $

The time-dependent tunneling rates are then defined as
\begin{align}
\label{eq:GammaP}
\Gamma_+ (t)
&=    \left| \xi \right|^2  \int_{-\infty}^t dt' \left[ \left\langle A_-(t) A_+ (t') \right\rangle + \left\langle A_-(t') A_+ (t) \right\rangle  \right]    \\
 \Gamma_- (t)
&=    \left| \xi \right|^2  \int_{-\infty}^t dt' \left[ \left\langle A_+(t) A_-(t') \right\rangle + \left\langle A_+(t') A_- (t) \right\rangle  \right]
  \label{eq:GammaM}
  \end{align}

Expanding the tunneling operators in terms of the anyon creation and annihilation operators, this further writes
\begin{align}
\Gamma_+ (t)
&=    \left| \xi \right|^2  \int_{-\infty}^t dt' \left[
\left\langle \Psi_1^\dagger (0,t)  \Psi_1 (0,t') \right\rangle
\left\langle  \Psi_2 (0,t) \Psi_2^\dagger (0,t')  \right\rangle \right. \nonumber \\
& \qquad \qquad \left.
+
\left\langle \Psi_1^\dagger (0,t') \Psi_1 (0,t)  \right\rangle
\left\langle \Psi_2 (0,t') \Psi_2^\dagger (0,t)  \right\rangle
 \right]    \\
 \Gamma_- (t)
&=    \left| \xi \right|^2  \int_{-\infty}^t dt' \left[
\left\langle  \Psi_1 (0,t)\Psi_1^\dagger (0,t')   \right\rangle
\left\langle \Psi_2^\dagger (0,t)\Psi_2 (0,t')  \right\rangle \right. \nonumber \\
& \qquad \qquad \left.
+
\left\langle \Psi_1 (0,t')\Psi_1^\dagger (0,t)   \right\rangle
\left\langle \Psi_2^\dagger (0,t')\Psi_2 (0,t)  \right\rangle
\right]
\end{align}
where one readily recognizes the correlation functions $\left\langle \Psi_i (t) \Psi_i^\dagger (t') \right\rangle$ and $\left\langle \Psi_i^\dagger (t) \Psi_i (t')  \right\rangle$, for each edge $i = 1,2$.

As the edge 2 is unperturbed, one can readily use the equilibrium result, so that
\begin{align}
\left\langle \Psi_2 (t) \Psi_2^\dagger (t') \right\rangle &= \left\langle \Psi_2^\dagger (t) \Psi_2 (t')  \right\rangle = \cG_\delta \left(  t-t' \right)
 \label{eq:G2}
\end{align}

The correlation function along the edge 1 is obtained by taking the thermal average over a prepared state corresponding to applying a voltage pulse, namely
\begin{align}
\left\langle \Psi_1 (t) \Psi_1^\dagger (t') \right\rangle &=
  \left\langle   \Psi_N (x_0, 0) \Psi_1 (0,t) \Psi_1^\dagger (0,t')   \Psi_N^\dagger (x_0,0) \right\rangle_0  \label{eq:corr}
  \end{align}
where $\Psi_N^\dagger$ was defined in Eq.~\eqref{eq:dis}. Using Eq.~\eqref{eq:comrelPsi} to work out explicitly the commutation relation between anyon current pulses and topological anyons, one obtains
\begin{align}
\left\langle \Psi_1 (t) \Psi_1^\dagger (t') \right\rangle
  &=
  e^{i \frac{\pi }{e} \int_{-\infty}^{+\infty} d\tau I_{N}(\tau) \left[ \text{Sign}(t'-\tau) - \text{Sign}( t-\tau ) \right]}
 \left\langle \Psi_1 (t) \Psi_1^\dagger (t') \right\rangle_0   \nonumber \\
    &=
 e^{- 2 i \frac{\pi }{e} \int_{t'}^{t} d\tau I_{N}(\tau) }
  \cG_\delta \left(  t-t' \right)   \nonumber \\
      &=
 e^{- i \theta \times N \left(t', t \right) }
 \cG_\delta \left(  t-t' \right)
   \label{eq:G1}
\end{align}
where we recover the braiding phase $\theta = 2 \pi \nu$ and we introduced the number of emitted anyons $N \left( t', t\right)$ crossing the QPC betwen times $t'$ and $t$, with $N \left( t', t\right) = \frac{1}{e^*} \int _{t'}^t d\tau I_N (\tau)$.

Substituting the expressions for the correlation functions Eqs.~\eqref{eq:G2} and \eqref{eq:G1} back into Eqs.~\eqref{eq:GammaP}-\eqref{eq:GammaM}, one is left with
\begin{align}
\Gamma_+ (t)
&=    \left| \xi \right|^2  \int_{-\infty}^t dt' \left[
 e^{ i \theta \times N \left(t', t \right) }
 \cG_\delta ^2\left(  t-t' \right)
+
 e^{ i \theta \times N \left(t, t' \right) }
 \cG_\delta ^2\left(  t'-t \right)
 \right]    \\
 \Gamma_- (t)
&=    \left| \xi \right|^2  \int_{-\infty}^t dt' \left[
e^{- i \theta \times N \left(t', t \right) }
 \cG_\delta ^2\left(  t-t' \right)
+
e^{- i \theta \times N \left(t, t' \right) }
 \cG_\delta^2 \left(  t'-t \right)
\right]
\end{align}
Using then that the conjugated equilibrium correlator satisfies $\left[ \cG_\delta \left(  t-t' \right) \right]^* = \cG_\delta \left(  t'-t \right)$, and that $N(t,t') = -N(t',t)$ is real, this finally reduces to
\begin{align}
\Gamma_\pm (t)
&=  2  \left| \xi \right|^2  \Re \left[ \int_{-\infty}^t dt'
 e^{ \pm i \theta \times N \left(t', t \right) }  \cG_\delta^2 \left(  t-t' \right)
 \right]  \label{eq:Gammpm}
\end{align}

\begin{figure}[h!]
    \centering
    \includegraphics[scale=0.8]{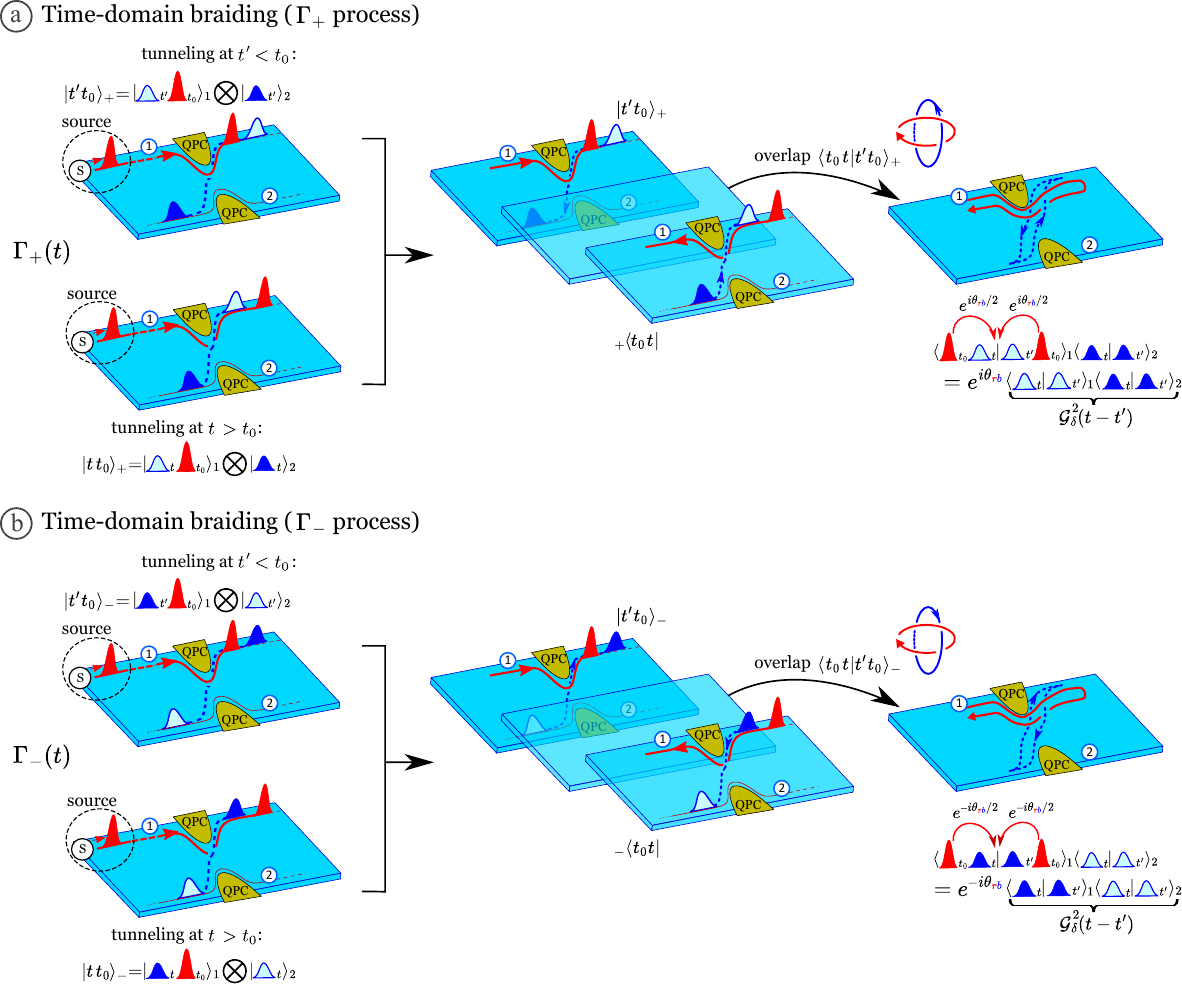}
   \legend{ \small{\textbf{Fig.S2: Anyon braiding at a QPC.} (\textbf{a}) Anyon braiding in the presence of an anyon source at input 1, process $\Gamma_+(t)$.  $|t t_0 \rangle_+$ and $|t' t_0 \rangle_+$ (left panel) represent the tunneling of the blue anyon from channel 1 to channel 2 after (for $t\geq t_0$) and before (for $t' \leq t_0$) the red anyon has crossed the QPC. The overlap  $\langle t_0 t|t' t_0 \rangle_+$ is sketched on the right panel, where a backward propagation in time is used to represent the ket $\langle t_0 t|$.  $\langle t_0 t|t' t_0 \rangle_+$ can be seen as a braiding process between the blue and red anyons. It differs from the overlap $\langle t|t' \rangle_+$ at equilibrium by the braiding phase factor $e^{i \theta_{rb}}$, where $ \theta_{rb}$ is the mutual braiding phase between the red and blue anyons.  (\textbf{b}) Anyon braiding in the presence of an anyon source at input 1, process $\Gamma_-(t)$.    $\langle t_0 t|t' t_0 \rangle_-$  can also be seen as a braiding between the blue and red anyons. However,  braiding occurs in the opposite direction compared to  $\langle t_0 t|t' t_0 \rangle_+$, such that the accumulated braiding phase factor is $e^{-i \theta_{rb}}$. This braiding phase factor can also be explicitly computed from the commutation relation between the anyon and anyon pulse creation operators derived in Eq.(\ref{eq:commutation}). This is illustrated by the sketches of $\langle t_0 t|t' t_0 \rangle_+$ and $\langle t_0 t|t' t_0 \rangle_-$ where the red arrows represent the phase $e^{\pm i \theta_{rb}/2}$ acquired from the exchange between an anyon and an anyon pulse operator. } \label{fig:timedomain} }    
\end{figure}

\section{Tunneling rates and the time-domain interferometry picture}

In the manuscript's text, we discuss the expression (\ref{eq:Gammpm}) of the tunneling rates in the context of a time-domain interferometry picture that was initially introduced in refs.\cite{Morel2022,Lee2022,Mora2022,Schiller2023}. The notation $| t' t_0\rangle_+$ and $| t' t_0\rangle_-$ introduced in the manuscript to describe the quantum states associated to forward and backward tunneling closely follow the notations introduced in Ref.\cite{Schiller2023}. These two quantum states can be directly connected to the anyon and anyon pulse creation and annihilation operators discussed above: 
\begin{align}
| t' t_0\rangle_+ &=  \Psi_2^\dagger (0,t') \Psi_1 (0,t')    \Psi_N^\dagger (x_0,0) |0 \rangle \\
| t' t_0\rangle_- &=   \Psi_1^\dagger (0,t') \Psi_2 (0,t')  \Psi_N^\dagger (x_0,0) |0 \rangle ,
\end{align} 
where $|0\rangle$ is the ground state. The product of the non-equilibrium correlation functions introduced above for the calculation of $\Gamma_+(t)$ and $\Gamma_-(t)$ can then be rewritten as:
\begin{align}
\left\langle \Psi_1^\dagger (t) \Psi_1 (t') \right\rangle \left\langle \Psi_2 (t) \Psi_2^\dagger (t') \right\rangle&=  \left \langle t_0 t|t' t_0\right \rangle_+  \\ 
\left\langle \Psi_1 (t) \Psi_1^\dagger (t') \right\rangle \left\langle \Psi_2^\dagger (t) \Psi_2 (t') \right\rangle&=   \left \langle t_0 t|t' t_0\right \rangle_-
\end{align}
and the tunneling rates as: 
\begin{align}
\Gamma_{\pm}(t) & = 2 \left| \xi \right|^2 \Re \left[ \int_{-\infty}^t dt'  \left \langle t_0 t|t' t_0\right \rangle_\pm \right].
\end{align}
This makes the connection between the anyon and anyon pulse creation and annihilation operators discussed in the supplementary and the bra and ket states discussed in the main manuscript. As discussed in Ref.\cite{Schiller2023}, the bra and ket notation is useful to emphasize the two interfering path in a time-domain interferometry picture.  Indeed, the tunneling current can be rewritten as:
\begin{align}
I_T(t) & = e^* \left [\Gamma_+(t) - \Gamma_-(t) \right] = e^*\left| \xi \right|^2  \left[ \int_{-\infty}^t dt'  \left( \left||t t_0 \rangle_+  + |t' t_0 \rangle_+|^2 - ||t t_0 \rangle_-  + |t' t_0 \rangle_-\right|^2  \right) \right],
\end{align}
showing that $\Gamma_+(t)$ (respectively $\Gamma_-(t)$) results from the interferometry between the paths  $|t t_0 \rangle_+$  and $|t' t_0 \rangle_+$ (resp. between the paths $|t t_0 \rangle_-$ and $|t' t_0 \rangle_-$) describing tunnel events occurring at different times. Fig.S2 sketches the tunneling processes $\Gamma_+(t)$ and  $\Gamma_-(t)$ (whereas due to space constraints, only $\Gamma_+(t)$ is discussed in the manuscript's text). On Fig.S2, the bra $|t' t_0 \rangle$ has a forward propagation in time while the time propagates backward for the ket  $\langle t t_0|$. This illustrates graphically that  the overlaps $\langle t t_0|t' t_0 \rangle_+$ and  $\langle t t_0|t' t_0 \rangle_-$ can be represented as a braiding between the blue and red anyons occurring in opposite directions. They thus differ from the equilibrium term by the phase factors $e^{i \theta_{rb}}$  and $e^{-i \theta_{rb}}$.

\section{Comparison between data and theory}

Comparisons between data and theory shown in the paper are performed by numerically computing Eqs.~(3) and (4) from the main text in the case where  periodic current pulses are generated. In this case, the number of anyons $N(t'=-\infty,t)$ emitted between time $t'=-\infty$ and time $t$ is a periodic function of $t$. The difference of the number of emitted anyons, $\Delta N(t'=-\infty,t)$, is also a periodic function of $t$ that also depends on the time delay $D$ between the two emitted anyon pulses. We thus introduce the Fourier coefficients $p_k(D)$ defined as:
\begin{eqnarray}
e^{i \theta \Delta N(t'=-\infty,t)} & =& \sum_k p_k(D) e^{i k \Omega t},
\end{eqnarray}
with $\Omega=2\pi f$. The Fourier coefficients thus depend on the shape of the generated current pulses, on the number $N$ of anyons carried per pulse, on the time delay between two pulses and on the braiding phase $\theta$.

The Fourier coefficients are numerically computed using a square shape of the pulse (see \ref{sec:fsize}) of temporal width $W$ and the value $\theta=2\pi/3$ of the braiding phase at $\nu=1/3$.  Once the Fourier coefficients are known,
 Eqs.(3) and (4) can be evaluated when periodic drives are applied:
\begin{eqnarray}
\Delta G_T & =& -2 C \frac{(e^*)^2}{\hbar} \sum_k P_k(D) \int_{0}^{+\infty} d\tau \big[\cos{(k \Omega \tau )} -1 \big] \Im \big[  \tau \mathcal{G}_{\delta}^2(\tau)\big] \\
\Delta S_T(D) & =& 4 C (e^*)^2  \sum_k P_k(D)  \int_{0}^{+\infty} d\tau \big[\cos{(k \Omega \tau )} -1 \big]  \Re \big[  \mathcal{G}_{\delta}^2(\tau)\big],
\end{eqnarray}
where $P_k(D)=|p_k(D)|^2$. Using $\big[\mathcal{G}_{\delta}(\tau)\big]^2 \propto \frac{1}{\left[ \sinh{\left(i \pi \frac{\tau_0 +i \tau}{\tau_{\text{th}}} \right) }\right]^{2\delta}}$, with $\tau_{\text{th}} = \hbar/(k_B T_{\text{el}})$, we obtain the following expressions for the normalized ratios $g(D)$ and $s(D)$:
\begin{eqnarray}
g(D) & =& \frac{\sum_k P_k(D) \Big(|\Gamma(\delta + i \frac{k \Omega \tau_{\text{th}}}{2\pi})|^2 \big[ \cosh{(\frac{k\Omega \tau_{\text{th}}}{2})} -\frac{2}{\pi} \Im[\Psi(\delta + i \frac{k \Omega \tau_{\text{th}}}{2\pi})]\sinh{(\frac{k\Omega \tau_{\text{th}}}{2})} \big] -\Gamma(\delta)^2 \Big)}{\sum_k P_k \left(D=\frac{1}{2f} \right) \Big( |\Gamma(\delta + i \frac{k \Omega \tau_{\text{th}}}{2\pi})|^2  \big[ \cosh{(\frac{k\Omega \tau_{\text{th}}}{2})} -\frac{2}{\pi} \Im[\Psi(\delta + i \frac{k \Omega \tau_{\text{th}}}{2\pi})]\sinh{(\frac{k\Omega \tau_{\text{th}}}{2})} \big] -\Gamma(\delta)^2 \Big)} \nonumber \\
\\
s(D) & =& \frac{\sum_k P_k(D) \Big( |\Gamma(\delta + i \frac{k \Omega \tau_{\text{th}}}{2\pi})|^2  \Im[\Psi(\delta + i \frac{k \Omega \tau_{\text{th}}}{2\pi})]\sinh{(\frac{k\Omega \tau_{\text{th}}}{2})}   \Big)}{\sum_k P_k \left(D=\frac{1}{2f} \right) \Big( |\Gamma(\delta + i \frac{k \Omega \tau_{\text{th}}}{2\pi})|^2  \Im[\Psi(\delta + i \frac{k \Omega \tau_{\text{th}}}{2\pi })]\sinh{(\frac{k\Omega \tau_{\text{th}}}{2 })}  \Big)},
\end{eqnarray}
where  $\Gamma$ and $\Psi$ are the Gamma and Digamma functions.

\section{Sample: Fractional quantum Hall effect and fractional charges $e^*$ at $\nu=1/3$}
\label{introduction}

This work studies the $\nu = 1/3$ fractional quantum Hall state.  A full sweep in magnetic field of the sample is presented in Fig.S3.a, showing the existence of quantum Hall states as vanishing or pronounced dips (for the less well-defined fractional states) in the backscattering probability $P$. The backscattering probability is defined as the ratio of the current backscattered at the QPC $I_T=I_3$ divided by the input current $I_0$ generated on input 2 (see inset of Fig.S3.a), $P=I_3/I_0=V_3/V_0$. For the measurement of $P$ presented on Fig.S3.a, the QPC gates are energized with a positive voltage of 50 mV, in order to reach a full suppression of the backscattering probability for well developed integer and fractional quantum Hall states (all the current $I_0$ flowing toward output 4). Integer quantum Hall states are visible for $B<5$ T and notable fractional states present are $\nu=2/3$, $\nu=2/5$, and $\nu=1/3$ for which the backscattering goes to zero. The experiments were performed on the $\nu=1/3$ state between $B=13.6$ T and $B=13.9 T$.  Once the magnetic field is set on the $\nu=1/3$ state, $P$ can be varied at will between an open ($P=0$) or closed ($P=1$) QPC by tuning the gate voltage. Shot noise measurements of the fractional charge of anyons at $\nu=1/3$ for different values of $P$ are presented on Fig.S3.b. A dc bias $V_0$ generates an input current $I_0 = V_0/(3h/e^2)$ toward the QPC. The excess current correlations $\Delta S_{4}$ are measured at output  4 and are expected to follow:

\begin{equation}
    \frac{\Delta S_{4}}{P(1-P)} = 2e^*I_0 \left(\coth(\frac{e^*V_0}{2k_BT_{el}})-\frac{2k_BT_{el}}{e^*V_0} \right) \label{eq:shotnoise}
\end{equation}
with $e^*=e/3$ for anyons at $\nu=1/3$, $T_{el}=40$ mK, and $P$ is monitored during the experiment (see Fig.S3.c). We observe $e^*=e/3$ charges for backscattering probabilities of the QPC set in the weak backscattering regime.

\begin{figure}[h]
    \centering
    \includegraphics[scale=0.5]{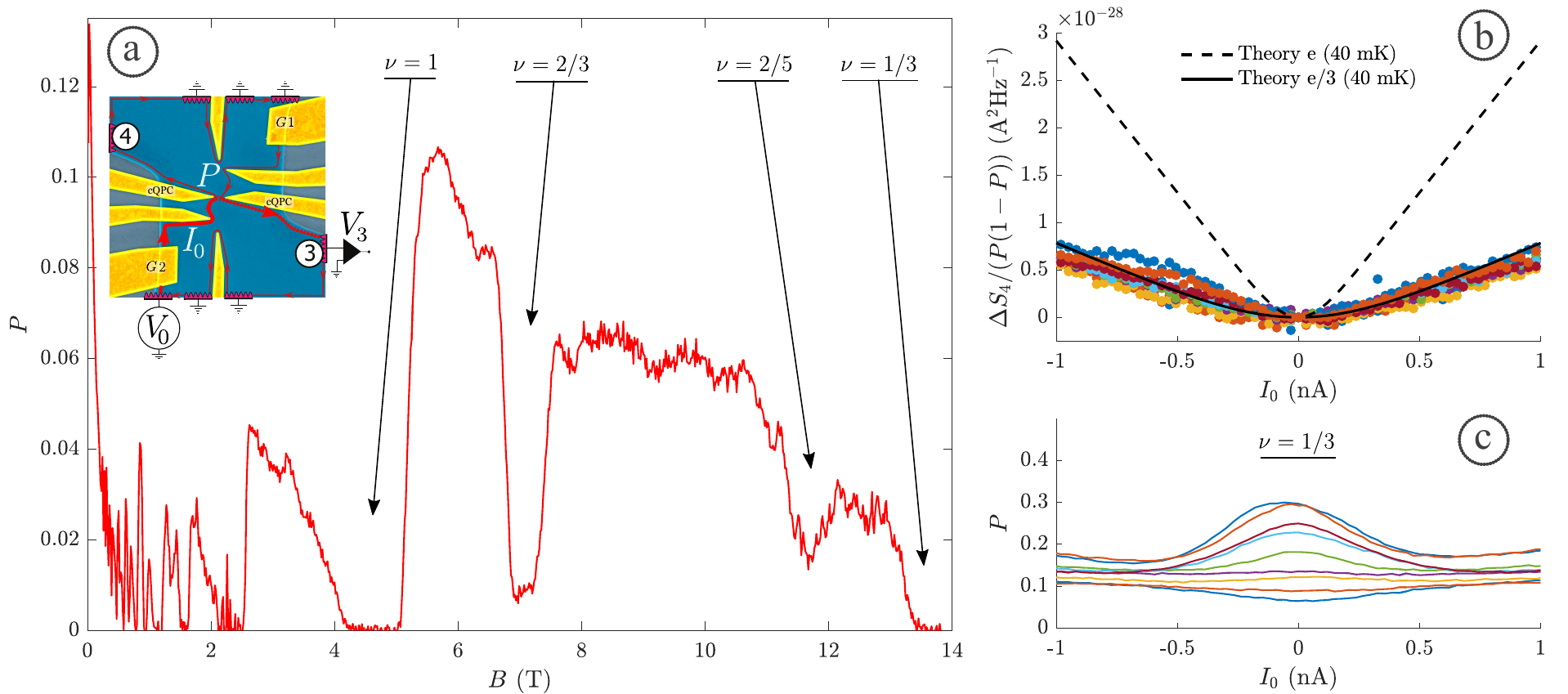}
    \legend{\small{\textbf{Fig.S3: Quantum Hall trace and shot noise at $B = 13.6 \;T$.}  \textbf{(a)} Magnetic field sweep of the sample used in the experiments. Vanishing or dips of $P=I_3/I_0$ as a function of $B$ denote the position of integer and fractional quantum Hall states, with a clear vanishing $P$ for $B>13.4$ T corresponding to the $\nu=1/3$ filling factor. Inset: schematic of the measurement setup for this quantum Hall trace. \textbf{(b)} Fractional charges measurement by a shot noise experiment. The current noise $\Delta S_{4}/(P(1-P))$ after the QPC is plotted as a function of the current input $I_0$ for different backscattering probabilities of the QPC, set in the weak backscattering regime. \textbf{(c)} Monitoring of the QPC backscattering probability $P$ as a function of $I_0$ during the shot noise experiment. The different traces correspond to different gate voltage settings of the QPC.}}
    \label{fig:sweep}
\end{figure}

\section{Additional data}

\begin{figure}
\centering
\includegraphics[width=0.75
\columnwidth,keepaspectratio]{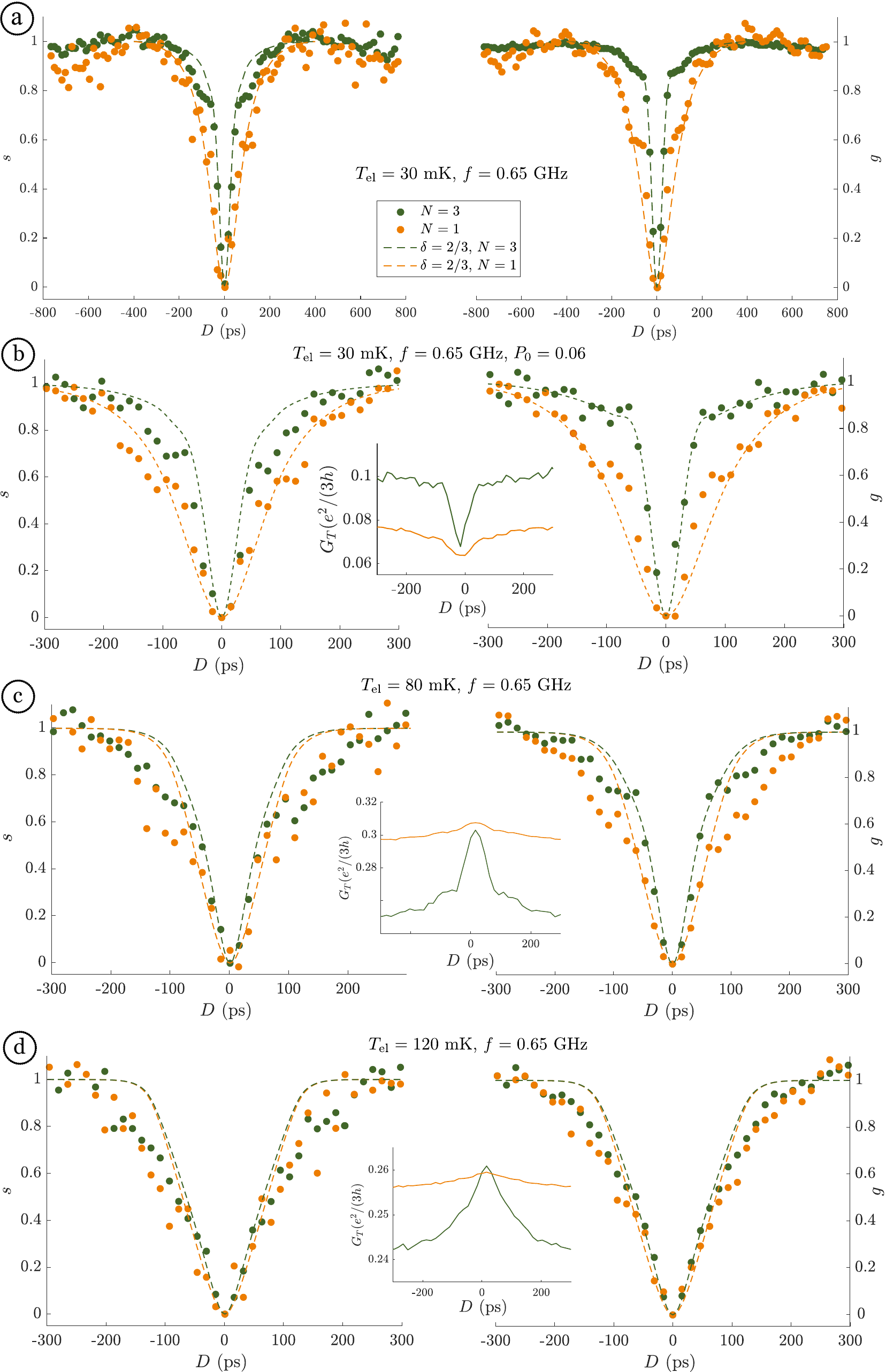}
\legend{ \small{\textbf{Fig.S4: HOM traces, additional data for $f=0.65$ GHz.}  (\textbf{a}) HOM traces for the full range of $|D| \leq \frac{1}{2f}$, $T_{\text{el}}=30$ mK. (\textbf{b}) HOM traces for $T_{0}=0.06$, $T_{\text{el}}=30$ mK. The green and orange dashed lines are numerical calculations with $T_{\text{el}} = 30$ mK, $W=0.065/f=100$ ps, $\delta=2/3$. The inset shows the measurements of the differential conductance $G_T(D)$. (\textbf{c}) HOM traces for $T_{\text{el}}=80$ mK. The green and orange dashed lines are numerical calculations with $T_{\text{el}} = 80$ mK, $W=0.065/f=100$ ps, $\delta=2/3$.(\textbf{d}) HOM traces for $T_{\text{el}}=120$ mK. The green and orange dashed lines are numerical calculations with $T_{\text{el}} = 120$ mK, $W=0.083/f=128$ ps, $\delta=2/3$. \label{Fig:S1}}}
\end{figure}

\subsection{HOM traces for $-1/(2f) \leq D \leq 1/(2f)$, $f=0.65$ GHz.}

We present on Fig.S4.a the HOM traces of $s$ and $g$ in the full scale of time delays $D$ spanning one full period of the excitation drive, $1/f=1.54$ ns (Fig.4 of the manuscript shows a zoom on short times $|D| \leq 300$ ps). Although the noise $s$ is almost flat for large time delays, $s( |D| \geq 300 ps) \approx 1)$, a small decrease of $s$ is observed for $|D| \geq 500$ ps. As a consequence, the normalization of $s$ is performed for $320 ps \leq |D| \leq 480 ps$.

\subsection{HOM traces for $P_0=0.06$, $f=0.65$ GHz.}

We present on Fig.S4.b the HOM traces measured in the very weak backscattering regime, $P_0=0.06$. As can be seen on the inset of Fig.S4.b, the evolution of $G_T$ when varying the time delay is opposite to the predictions of the Luttinger model: $G_T$ increases when increasing the delay $D$ (which corresponds to an increasing value to the applied time-dependent voltage averaged on one period of the excitation drive). Although an increase of $G_T$ with $D$ would suggest $\delta >1$ (corresponding to an attractive interaction between electrons), the measurement of the width of the HOM traces agrees with $\delta \approx 2/3$, consistently with other values of $P_0$ shown in the main manuscript.

\subsection{HOM traces for $T_{\text{el}}=80$ mK and $T_{\text{el}}=120$ mK, $f=0.65$ GHz.}

We present on Fig.S4.c and Fig.S4.d the HOM traces measured for the highest temperatures $T_{\text{el}}=80$ mK and $T_{\text{el}}=120$ mK. At these higher temperatures, the signal over noise ratio is much weaker, as both the noise signals $\Delta S_{4}$ and the non-linearities of $G_T$ are almost suppressed. However, the expected effect of further increasing the temperature is observed. For $T_{\text{el}}=80$ mK (see Fig.S4.c), a small increase of the width between $N =3$ and $N=1$ is observed on the normalized differential conductance $g$. Despite the poor signal over noise ratio for the noise measurements at this temperature, this small increase of the width can also be seen on the normalized noise $s$. At the highest temperature of $T_{\text{el}}=120$ mK (see Fig.S4.d), the difference between $N=3$ and $N=1$ can barely be observed, both on the measurements of $s$ and $g$. The agreement with theory (green and orange dashed lines) is poor at $T_{\text{el}}=80$ mK, due to a widening of the tails of the HOM traces (for $|D| \geq 100$ ps) observed both for $N =3$ and $N=1$ which is not captured by the model. However, the measured excess width $\Delta D_{1/2}=20\pm 6$ ps, which is less sensitive to the specific shape of the HOM traces, agrees with the prediction for $\delta =0.66 \pm 0.08$ at $T_{\text{el}}=80$ mK within error bars (the prediction for  $T_{\text{el}}=80$ mK and $\delta =0.58$ is $\Delta D_{1/2}=20$ ps).

\subsection{HOM trace for $f=0.5$ GHz.}
The shape of the excitation drive and of the emitted anyon current pulses depend crucially on the excitation frequency. It is particularly affected by spurious resonances in the transmission of the high frequency coaxial lines that propagate the excitation voltage through the fridge from the room temperature to the sample. The frequency $f=0.65$ GHz is chosen to minimize the deformation of the emitted current pulses by these spurious effects.  Fig.S5 presents the measurement of $s(D)$ at $T_{\text{el}}=30$ mK and at a different frequency $f=0.5$ GHz. As can be seen on the figure, the HOM dips of $s$ are wider compared to the case of $f=0.65$ GHz discussed so far. This is related to a widening of the applied voltage pulse during the propagation in the fridge. However, the measured width differences $\Delta D_{1/2}$ between $N=3$ and $N=1$ almost agree with each other within error bars: we obtain $\Delta D_{1/2} =50 \pm 5 $ ps for $f=0.5$ GHz, compared to $\Delta D_{1/2} =39 \pm 4.5 $ ps for $f=0.65$ GHz.

\begin{figure}
\centering
\includegraphics[width=0.5
\columnwidth,keepaspectratio]{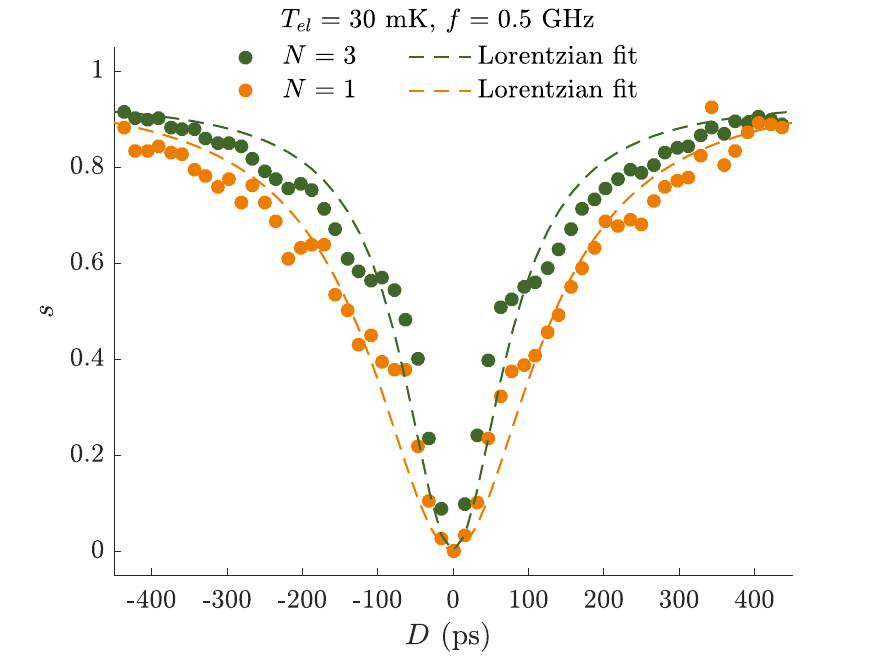}
\legend{ \small{\textbf{Fig.S5: HOM trace $s(D)$ for $T_{\text{el}}=30$ mK and $f=0.5$ GHz.} The green and orange dashed lines are Lorentzian fits of $s(D)$ from which we extract the width difference $\Delta D_{1/2}=50 \pm 5$ ps.\label{Fig:S2}}}
\end{figure}

\section{Role of the temporal width of the anyon pulses on the determination of the braiding phase}

As discussed in the manuscript, the non-trivial braiding between the incoming anyon pulses and the anyon excitations tunneling at a QPC can be evidenced by the widening of the HOM dips of the current noise and of the differential conductance. The suppression of braiding effects is thus characterized by the observation of a minimal width of the HOM dip for a given value $N_{min}$ of the number of anyons carried per pulse for which the braiding phase is trivial: $\theta \times N_{min} =2\pi$. The anyon braiding phase is then directly deduced from $\theta=2\pi/N_{min}$. Fig.5 of the manuscript presents the normalized excess width of the HOM dips $\Delta \Tilde{D}_{1/2}$ as a function of $N$. We observe a smooth transition from $\Delta \Tilde{D}_{1/2}=1$ for $N\leq 1$ to $\Delta \Tilde{D}_{1/2}=0$ for $N=3$. Interestingly, the transition may be much sharper by using anyon pulses with a smaller temporal width. Additional simulations of $\Delta \Tilde{D}_{1/2}(N)$ for different values of $W$ are presented in this supplementary part on Fig.S6. As can be seen on the figure, by decreasing the pulses temporal width, the minimum of $\Delta \Tilde{D}_{1/2}(N)$ for $N=N_{min}$ becomes sharper. In particular, an increase of $\Delta \Tilde{D}_{1/2}(N)$ for $N>N_{min}$ may be observed for $W<70$ ps, which is expected as the mutual braiding phase becomes again non-trivial ($\theta \times N >2\pi)$ for $N>3$. The observation of a flat value $\Delta \Tilde{D}_{1/2}(N)\approx 0$ for $N>3$ in Fig.5 of the manuscript can be explained by the width of the anyon pulses used in the experiment, $W=100$ ps. When using extremely sharp pulses with $W=15$ ps, $\Delta \Tilde{D}_{1/2}$ varies more abruptly and exhibits a sharp minimum for $N=3$. Therefore, using narrower incident pulses would allow us to be more precise in the measurement of $\theta$. However, generating such narrow current pulses is experimentally challenging. In conclusion, the width of the pulses is thus doubly important in the experiment, on a first level to have sharp enough pulses so that $W$ reaches the order of magnitude of $\tau_\delta$, and on a second level because improving on the sharpness of the pulses means improving on the precision of the determination of the braiding phase.

\begin{figure}[h]
    \centering
    \includegraphics[scale=0.4]{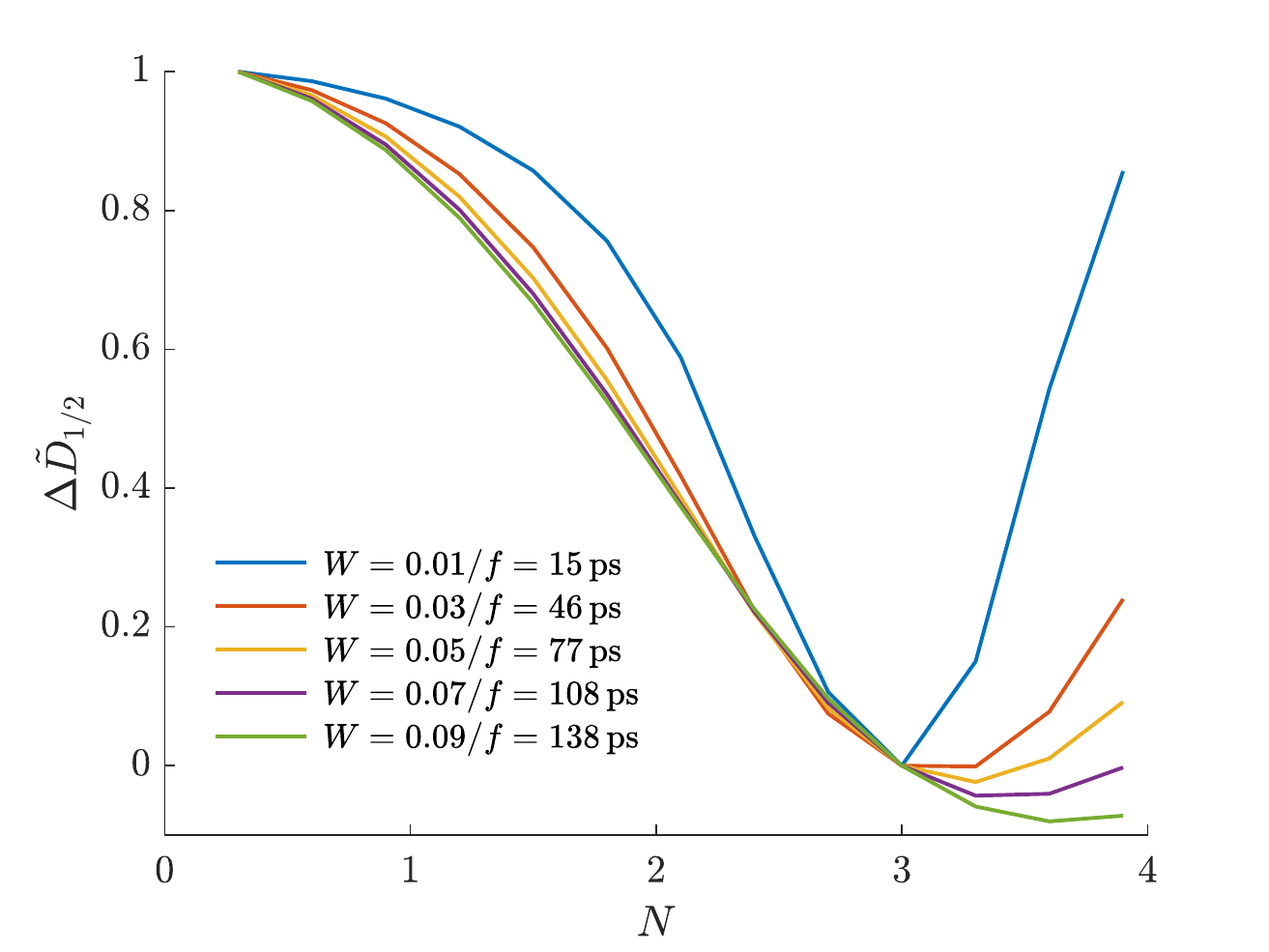}
    \legend{\small{\textbf{Fig.S6: $\Delta \Tilde{D}_{1/2}$ as a function of $N$.} Numerical simulations of the normalized excess width of the HOM dip $\Delta \Tilde{D}_{1/2} = \frac{D_{1/2}(N)-D_{1/2}(N=3)}{D_{1/2}(N \ll 1)-D_{1/2}(N=3)}$ as a function of $N$ for different values of the temporal width of the incoming pulses $W$ (15 ps, 46 ps, 77 ps, 108 ps and 138 ps). An important effect of braiding is recovered for $N>3$ ($\theta \times N > 2\pi$) only for the sharpest pulses, $W=15$ ps. The minimum at $N=3$ for $W=15ps$, denoting a trivial braiding such that $\theta \times N = 2\pi$, is less pronounced as the width of the pulses increases. The simulation parameters are $\theta = 2\pi/3$, $\delta = 0.66$, $T_{el} = 35$ mK.}}
    \label{fig:role_of_W}
\end{figure}

\end{document}